\documentclass[12pt,a4paper,hypertex]{article}
\usepackage{jheppub}

\allowdisplaybreaks[1]
\usepackage{amsmath,bm}
\usepackage{subfigure}

\title{Out-of-time-order correlators in quantum mechanics}

\author[1]{Koji Hashimoto,}
\author[2]{Keiju Murata}
\author[2]{and  Ryosuke Yoshii}
\affiliation[1]{Department of Physics, Osaka University, Toyonaka, Osaka 560-0043, Japan}
\affiliation[2]{Keio University, 4-1-1 Hiyoshi, Yokohama 223-8521, Japan}
\emailAdd{koji@phys.sci.osaka-u.ac.jp}
\emailAdd{keiju@phys-h.keio.ac.jp}
\emailAdd{ryoshii@phys.keio.ac.jp}

\abstract{%
The out-of-time-order correlator (OTOC) is considered as a measure of quantum chaos.
We formulate how to calculate the OTOC for quantum mechanics with a general Hamiltonian.
We demonstrate explicit calculations of OTOCs for a harmonic oscillator,
a particle in a one-dimensional box, a circle billiard and stadium billiards.
For the first two cases, OTOCs are periodic in time because of their commensurable energy spectra.
For the circle and stadium billiards, they are not recursive but saturate to constant values
which are linear in temperature.
Although the stadium billiard is a typical example of the classical chaos,
an expected exponential growth of the OTOC is not found.
We also discuss the classical limit of the OTOC.
Analysis of a time evolution of a wavepacket in a box shows that the OTOC can deviate
from its classical value at a time much earlier than the Ehrenfest time. 
}

\preprint{OU-HET-926}

\begin{document}

\maketitle

\section{Introduction and summary}
\label{intro}

The out-of-time-order correlator (OTOC) is typically defined by 
\begin{align}
C_T(t) \equiv -\langle  [W(t),V(0)]^2 \rangle\ ,
\label{OTOCWV}
\end{align}
where $\langle\cdots\rangle$ represents the thermal average.
$W(t)$ and $V(t)$ are operators at time $t$ in the Heisenberg representation.
The OTOC, first introduced in a calculation of a vertex correction of a current for a superconductor~\cite{Larkin}, 
was recently turned to be considered as a measure of the magnitude of quantum chaos. 
A naive argument for the relation between the OTOC and chaos  
is as follows~\cite{Maldacena:2015waa}.
Consider position and momentum operators, $x(t)$ and $p(t)$, in a quantum system.
We can define an OTOC as $C_T=-\langle [x(t),p(0)]^2 \rangle$. 
Taking a naive semiclassical limit, 
we would be able to replace the commutator $[x(t),p(0)]$ by the Poisson bracket 
$i\hbar \{x(t),p(0)\}_\textrm{PB}=i\hbar \delta x(t)/\delta x(0)$.
For a classically chaotic system with a Lyapunov exponent $\lambda$,
we have $\delta x(t)/\delta x(0)\sim e^{\lambda t}$ because of the sensitivity to initial conditions. 
Thus, the OTOC should grow as $\sim \hbar^2 e^{2\lambda t}$ and
we can read off the quantum Lyapunov exponent $\lambda$ from it.
The quantization of a classically chaotic system 
may provide a positive quantum Lyapunov exponent of the OTOC.
A possible distinction from the classical chaotic system is that
the OTOC does not grow eternally but saturates at the Ehrenfest time $t_E$.
The Ehrenfest time is defined by the time scale beyond which the wave function spreads over the whole system. 
It is roughly 
characterized as a boundary between a particle-like behavior and a wave-like behavior of
the wave function.

In recent years,
the OTOC has been regarded as an important observable
in the context of AdS/CFT correspondence~\cite{Maldacena:1997re} or quantum gravity.
A maximum bound of the quantum Lyapunov exponent was proposed as
$\lambda\leq 2\pi k_{\rm B} T/\hbar$~\cite{Maldacena:2015waa}. 
The bound was originally suggested in the context of quantum information
around black hole horizons~\cite{Shenker:2013pqa,Shenker:2013yza}
(see also Refs.\cite{Leichenauer:2014nxa,Kitaev-talk,Shenker:2014cwa,Jackson:2014nla,Polchinski:2015cea}).
The Lyapunov bound is saturated by the Sachdev-Ye-Kitaev (SYK) model~\cite{Sachdev:1992fk,Kitaev-talk-KITP}:
A quantum mechanics of Majorana fermions with infinitely long range disorder interactions.
Saturation of the quantum Lyapunov bound indicates that 
the SYK model describes a quantum black hole through the AdS/CFT correspondence.

Since the original calculation of the quantum Lyapunov exponent by Kitaev,
there appeared subsequent study for generalizing the SYK model \cite{Gross,Witten,Terashima}.
However, we are still missing explicit examples of OTOCs.
Do typical chaotic systems show exponential growth in OTOCs?
Can we find any qualitative difference between integrable and chaotic systems through OTOCs?
To answer these problems, we study the OTOC of single particle quantum mechanics. 
First we formulate how to calculate the OTOC for generic quantum mechanics.
In particular, by the reason described above, we choose $W=x$ and $V=p$ to
measure a possible indication of quantum chaos.
Based on the formalism,
we 
examine OTOCs of some popular quantum systems:
(i)a harmonic oscillator,
(ii)a particle in a one-dimensional box,
(iii)a circle billiard (a particle in a 
circle-shaped infinite well),
and
(iv)a stadium billiard.
Former three are known as integrable systems.
The stadium billiard \cite{Sinai,Bunimovich1,Bunimovich2,Bunimovich3,Benettin}, on the other hand, 
is one of the most popular and well-studied Hamiltonian chaotic systems.\footnote{
Our targets are time-independent Hamiltonian systems where energy is conserved.
As an example of time-dependent Hamiltonian systems,
an OTOC for a kicked rotor system has been studied in Ref.\cite{Rozenbaum}.
}

\begin{figure}
  \centering
  \subfigure[Particle in a box]
  {\includegraphics[scale=0.45]{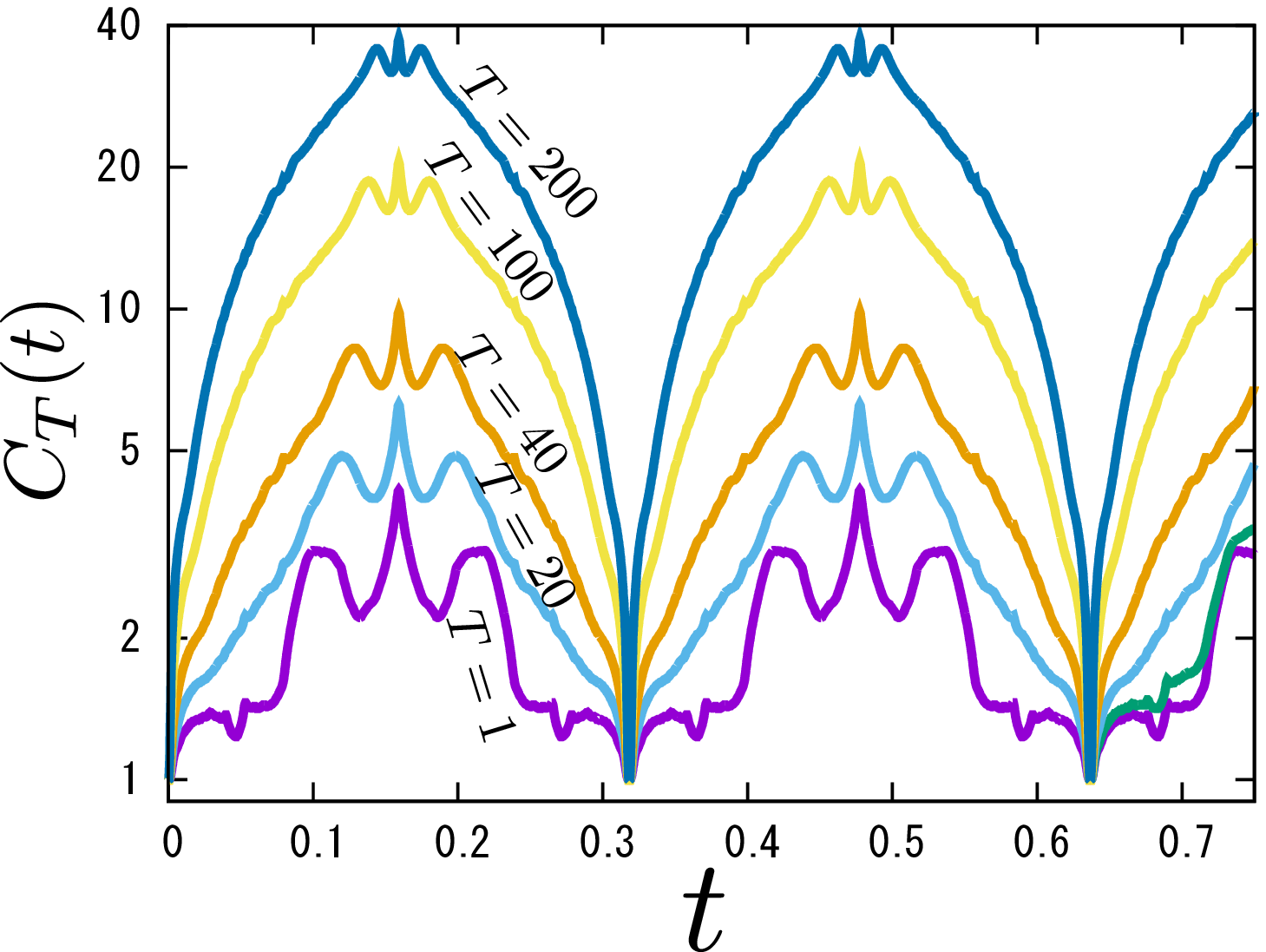}\label{otocbox2}
  }
  \subfigure[Stadium billiard]
 {\includegraphics[scale=0.45]{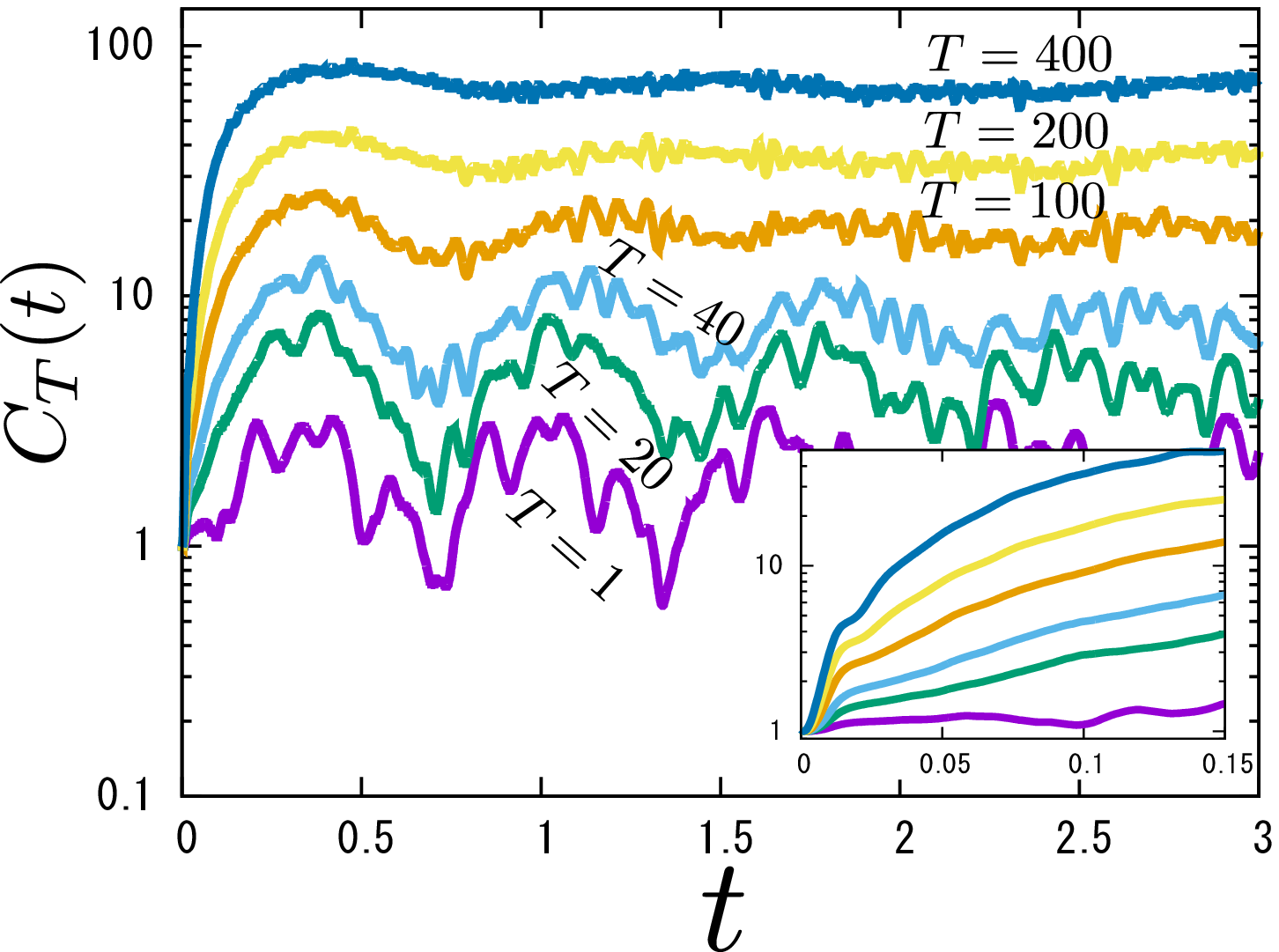}\label{th_otoc_long}
  }
 \caption{
The OTOC $C_T(t) = -\langle [x(t),p(0)]^2\rangle$ 
of a particle in a 1D box (a) and 
that of a billiard (b). 
$T$ represents the temperature of the systems. 
We find a clear distinction between the two: The OTOC for the particle in a box
periodically comes back to its initial value ($=1$), while
that for stadium billiard saturates to a constant value. 
 The asymptotic value grows with the temperature.
 In the inset of the right panel, we show the early time evolution of the OTOC.
 We find no clear exponential growth of the OTOC.
 }
 \label{Bil-Res}
\end{figure}

Among our main results, we show two numerical results in Fig.~\ref{Bil-Res},
which shows typical behavior of the OTOCs. 
The left panel is a numerical evaluation of the 
OTOC for a particle in a 1D box, and the 
right is that for a stadium billiard. 
In the figure, we took the unit of $\hbar=k_B=2m=1$ where $m$ is the mass of the particle. 
We also set $(\textrm{Length of the box})=1$ or $(\textrm{Area of the billiard})=1$.
We summarize our findings in this paper below:
\begin{enumerate}
\item The OTOCs grow at early times.
      However, at least for $T\lesssim 400$,
      they do not show apparent exponential growth even for the stadium billiards.
\item
The OTOC of a particle in a box is periodic in time because of its commensurable energy spectrum, while
     that of the stadium billiards saturates to a constant value.
 \item
      The OTOCs grow with temperature except for the harmonic oscillator.
       The high temperature limit does not reproduce the classical value in general.
 For high temperature,
       asymptotic or maximum values of OTOCs grow linearly in temperature as
       $C_T\sim m T\times (\textrm{typical system size})^2$.
\item
An analysis of a time evolution of a wave packet 
in a 1D box shows that the OTOC deviates from its classical value at
the time scale parametrically earlier than the Ehrenfest time $t_{\rm E}$.
\end{enumerate}

\subsubsection*{Organization of this paper}

We start in Sec.~\ref{intro} with an introduction and a summary of our results obtained
in this paper.
We formulate how to calculate the OTOCs 
in a general quantum mechanical system in Sec.~\ref{sec:OTOC}. Then in Sec.~\ref{sec:Intex} we evaluate
the OTOCs for integrable examples such as the harmonic oscillator, a particle in a 1D box,
and a particle in a circular billiard. In Sec.~\ref{sec:non-int}, 
after reviewing the classical chaos of the stadium billiards,  
we present our numerical results of the OTOCs for the quantum stadium billiards. 
In Sec.~\ref{sec:Ontheclass}, 
we study time evolution of a wave packet in a 1D box, to find a deviation of 
the OTOC from its classical value at rather early times.
Sec.~\ref{sec:Diss} is devoted for discussions, with a relation to
quantum fidelity and Loschmidt echo.
Our appendices include description
on our numerical truncation errors and
detailed formulas for the analytic calculation of the OTOC for a wave packet.

\subsubsection*{Units in this paper}
Throughout this paper, we work with 
the unit of $\hbar=k_B=2m=1$, where $m$ is the mass of a particle.
When we consider a particle in the 1D box and the stadium billiard,
we also set $(\textrm{length of the box})\equiv L=1$ and $(\textrm{area of the billiard})\equiv A=1$, respectively.
For the billiard, one can easily restore dimensional parameters notifying 
\[
 \text{Time}\sim\frac{2mA}{\hbar}\ ,\quad
  \text{Energy}\sim\frac{\hbar^2}{2mA}\ ,\quad
  \text{Length}\sim \sqrt{A}\ .
\]
For the particle in the box, $A$ is replaced by $L^2$.

\section{Out-of-time-order correlators in quantum mechanics}
\label{sec:OTOC}

In this section, 
we propose a formalism to compute the OTOC for general quantum mechanics with 
time-independent Hamiltonian: $H=H(x_1,\cdots,x_n,p_1,\cdots,p_n)$. 
We consider the out-of-time-order correlator (OTOC) defined by 
\begin{equation}
 C_T(t)=-\langle [x(t),p(0)]^2 \rangle\ , 
 \label{otoc_def}
\end{equation}
where $\langle \mathcal{O}\rangle \equiv \textrm{tr}[e^{-\beta H}\mathcal{O}]/\textrm{tr}e^{-\beta H}$.
Here we define $\beta=1/T$ with the temperature of the system $T$. 
We denoted $x=x_1$ and $p=p_1$ for notational simplicity.
Hereafter, we will omit the argument of Heisenberg operators for $t=0$: $\mathcal{O}\equiv \mathcal{O}(0)$.
Taking energy eigenstates as the basis of the Hilbert space, we can rewrite the OTOC as
\begin{equation}
 C_T(t)=\frac{1}{Z}\sum_n e^{-\beta E_n} c_n(t)\ ,\qquad c_n(t)\equiv -\langle n | [x(t),p]^2 |n \rangle \ ,
  \label{CT}
\end{equation}
where $H|n\rangle =E_n |n\rangle$.
We will refer the OTOC for a fixed energy eigenstate, $c_n(t)$, as a microcanonical OTOC.
On the other hand, we will refer $C_T(t)$ as a thermal OTOC.
Once we compute microcanonical OTOCs, we can obtain the thermal OTOC by taking their thermal average.\footnote{
{
In the definition of the thermal OTOC in Ref.~\cite{Maldacena:2015waa},
the fourth roots of the thermal density matrix, $y=(e^{-\beta H}/Z)^{1/4}$, are inserted between the operators.
In this paper, on the other hand,
we just take an ordinary thermal average as in Eq.~\eqref{CT}, 
which gives another natural definition of the thermal OTOC.}
}
Let us rewrite the microcanonical OTOC using matrix elements of $x$ and $p$ for numerical calculations.
Using the completeness relation $1=\sum_m |m\rangle \langle m|$,  we rewrite the microcanonical OTOC as
\begin{equation}
 c_n(t)=\sum_m b_{nm}(t)b^\ast_{nm}(t)\ ,\qquad b_{nm}(t)\equiv -i\langle n | [x(t),p] |m \rangle\ .
\label{cbb}
\end{equation}
Note that $b_{nm}(t)$ is Hermitian: $b_{nm}(t)=b_{mn}^\ast(t)$.
Substituting $x(t)=e^{iHt} x e^{-iHt}$ and inserting the completeness relation again, 
we obtain
\begin{equation}
 b_{nm}(t)=-i\sum_{k} (e^{iE_{nk}t}x_{nk}p_{km}-e^{iE_{km}t}p_{nk}x_{km})\ ,
\label{bnm1}
\end{equation}
where $E_{nm}=E_n-E_m$, $x_{nm}\equiv \langle n|x|m\rangle$ and $p_{nm}\equiv \langle n|p|m\rangle$.
In this expression, there are matrix components of $p$.
They are not desirable since numerical derivatives of wave functions lose the numerical accuracy.
For a natural Hamiltonian with the form, 
\begin{equation}
 H=\sum_{i=1}^N p_i^2+ U(x_1,\cdots,x_N)\ ,
 \label{nat_Ham}
\end{equation}
we can express $p_{nm}$ using $x_{nm}$. 
From Eq.(\ref{nat_Ham}), we have $[H,x] = -2ip$.
Applying $\langle m |\cdots |n\rangle $ to the both sides of the equation,
we obtain
\begin{equation}
 p_{mn}=\frac{i}{2}E_{mn}x_{mn}\ .
\end{equation}
Substituting this expression into Eq.(\ref{bnm1}), we have
\begin{equation}
 b_{nm}(t)=\frac{1}{2}\sum_{k} x_{nk}x_{km} (E_{km}e^{iE_{nk}t}-E_{nk}e^{iE_{km}t})\ .
\label{bnm2}
\end{equation}
Therefore, once we know the matrix elements of $x$ and the energy spectrum $E_n$,
we can compute OTOCs through Eqs.(\ref{bnm2}), (\ref{cbb}) and (\ref{CT}).

For actual numerical calculations, we need truncation for the summations 
in Eqs.(\ref{bnm2}), (\ref{cbb}) and (\ref{CT}).
In appendix.\ref{truncerror}, we check that our results do not depend on the truncation when we take 
the truncation cut-off sufficiently large.

\section{Integrable examples}
\label{sec:Intex}

\subsection{Harmonic oscillator}

For concreteness, we will show some explicit calculation for the OTOC in integrable systems.
One of the simplest integrable examples is the 1D harmonic oscillator, 
\begin{equation}
 H= p^2+\frac{\omega^2}{4}x^2\ .
\end{equation}
Although OTOCs for the harmonic oscillator have been already studied in Ref.\cite{Hashimoto:2016wme},
we compute them again using the formalism in the previous section.
The energy spectrum and matrix elements of $x$ is given by
\begin{equation}
 E_n=\left(n+\frac{1}{2}\right)\omega\ ,\quad
 x_{nm}=\frac{1}{\sqrt{\omega}}(\sqrt{m}\delta_{n,m-1}+\sqrt{m+1}\delta_{n,m+1})\ ,
\end{equation}
where $n,m=0,1,2,\cdots$.
Substituting above expressions into Eq.(\ref{bnm2}), we have
\begin{equation}
 b_{nm}(t)=\delta_{nm} \cos\omega t\ .
\end{equation}
Therefore, from Eqs.~(\ref{cbb}) and (\ref{CT}), we obtain OTOCs as
\begin{equation}
 c_n(t)=\cos^2 \omega t\ ,\qquad C_T(t)=\cos^2 \omega t\ .
 \label{harmOTOC}
\end{equation}
They are periodic functions whose periodicity is $\Delta t=\pi/\omega$. 
They do not depend on energy level $n$ or temperature $T$.
We will find that this is a special property only for the harmonic oscillator amongst our examples. 

As in Ref.\cite{Hashimoto:2016wme},
one can also get the same result using the explicit expression of the Heisenberg operators:
\begin{equation}
 x(t)=x(0)\cos \omega t + \frac{2}{\omega}p(0)\sin \omega t\ ,\qquad
  p(t)=p(0)\cos \omega t -\frac{\omega}{2}x(0)\sin \omega t\ .
 \label{xpsol_harm}
\end{equation}
From the explicit solution, we have $[x(t),p(0)]=i\cos \omega t$ and  OTOCs are given as Eq.(\ref{harmOTOC}).
This method is not useful for other cases since 
it is difficult (or impossible) to obtain explicit expressions
of Heisenberg operators for a general Hamiltonian.

\subsection{Particle in a box}

\begin{figure}
\begin{center}
\includegraphics[scale=0.5]{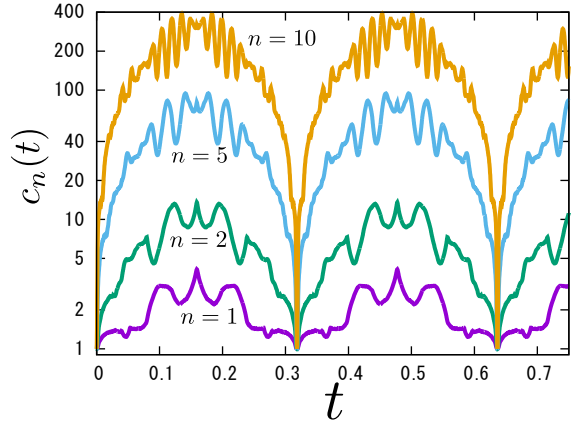}
\end{center}
\caption{
Microcanonical out-of-time-order correlators for the particle in a 1D box ($L=1$).
}
\label{motocbox}
\end{figure}

One of the other integrable examples is a particle in a 1D box. The Hamiltonian for the one-dimensional case
is
\begin{equation}
 H=p^2+V_\textrm{box}(x)\ ,\qquad V_\textrm{box}(x)=
 \begin{cases}
  0 & 0<x<1\\
  \infty & \textrm{else}
 \end{cases}\ .
\end{equation}
Eigenfunctions and eigenvalues are given by 
\begin{equation}
 \psi_n=\sqrt{2} \sin \pi n x\ , \qquad E_n=\pi^2 n^2\ ,
  \label{PBOX_efev}
\end{equation}
where $n=1,2,\cdots$.
The matrix elements of $x$ are written as
\begin{equation}
 x_{nm}=
 \begin{cases}
 \frac{1}{2} & (n= m)\\
 \frac{1-(-1)^{n+m}}{\pi^2}\left[\frac{1}{(n+m)^2}-\frac{1}{(n-m)^2}\right] & (n\neq m)
  \end{cases}\ .
  \label{xnm_pinb}
\end{equation}
Although we know exact expressions of $x$ and energy eigenvalues, 
it would be impossible to carry out the summation in Eq.(\ref{bnm2}) analytically.
So, we evaluate $b_{nm}(t)$ and $c_{n}(t)$ numerically with truncation $n,m\leq N_\textrm{trunc}=100$
and compute OTOCs. 
In Figs.\ref{motocbox} and \ref{otocbox2},
we show microcanonical and thermal OTOCs, respectively. 
Note that the energy spectrum for the particle in a box is commensurable: $E_n$ is proportional to the integer $n^2$.
By using Eq.~(\ref{xnm_pinb}), one can show that the all $E_{kl}$ appearing in rhs of Eq.~(\ref{bnm2}) become $\pi^2 \times$ odd integer and thus, OTOCs have periodicity $\Delta t=1/\pi$. 
For large $n$, microcanonical OTOCs become large and tend to oscillate since 
high frequency modes become relevant in Eq.(\ref{bnm2}).
High frequency oscillations seem to be suppressed in the thermal OTOC.
The thermal OTOC also tends to be large at high temperature. 
We have found
that the maximum of the thermal OTOC increases linearly as a function of $T$:
$\max C_T\simeq 0.1672\times 2mTL^2$ ($mTL^2\gg 1$) 
where the size of the box $L$ and the mass of the particle $m$ are restored.
We have also checked that the time average of the OTOC is given by 
\begin{equation}
 \bar{C}_T\simeq 0.0836 \times 2mTL^2\qquad (mTL^2\gg 1)\ ,
 \label{Cave_pinb}
\end{equation}
where $\bar{C}_T=\lim_{\tau\to\infty}\int^\tau_0 dt C_T(t)/\tau$.

For the particle in a 2D square box, $V(x,y)=0\ (0<x<1, 0<y<1)$, $\infty (\textrm{else})$,
we can obtain the same result for the thermal OTOC. 
Eigenstates in the 2D box are completely separable as $|n_x\rangle|n_y\rangle$,
so the operator $[x(t),p(0)]^2$ does not operate to $|n_y\rangle$.
Therefore, the existence of the $y$-direction is completely irrelevant for calculating the thermal OTOC.

\subsection{Circle billiard}

As a non-trivial 2D example, we consider a circle billiard:
\begin{equation}
 H=p_1^2+p_2^2+V_\textrm{circ}(x,y)\ ,\qquad V_\textrm{circ}(x,y)=
 \begin{cases}
  0 & x^2+y^2<R^2\\
  \infty & \textrm{else}
 \end{cases}\ .
\end{equation}
In this case, the $x$- and $y$-directions are not separable unlike the 2D square box. 
It is known that classical dynamics of a particle in the circle billiard is integrable. (We will see that in section.\ref{classical}.)
We fix the radius of the circle as $R=1/\sqrt{\pi}$ so that the area of the billiard becomes unity.
Taking polar coordinates $x=r\cos\theta$ and $y=r\sin\theta$,
we obtain exact expressions for eigenvalues and eigenfunctions as
\begin{equation}
 E_{kl}=\pi \rho_{kl}^2\ ,\qquad \psi_{kl}=\mathcal{N} J_k(\sqrt{\pi} \rho_{kl} r) e^{ik\theta}\ ,
\end{equation}
where $k\in \bm{Z}$ and $l\in \{1,2,\cdots\}$.
$J_k$ is the Bessel function of the first kind
and $\rho_{kl}$ represents its $l$-th root, i.e. $J_{k}(\rho_{kl})=0$.
The normalization factor is given by
$\mathcal{N}^{-1}=\pi J_{k+1}(\rho_{kl})$. 

The energy spectrum for the circle billiard is not commensurable.
It is only asymptotically commensurable:
It tends to be commensurable for high energy because of $\rho_{k l}\simeq (k/2+l)\pi$ for $l\gg 1$.
Although eigenvalues and functions are labeled by two integers $k$ and $l$, 
we relabel them by a single integer $n$ in ascending order of $E_{kl}$ and denote them as $(E_n,\psi_n)$.
The matrix elements of $x$ can be obtained from
\begin{equation}
 x_{nm}=\int^{1/\sqrt{\pi}}_0 rdr \int_0^{2\pi}d\theta\, \psi_n^\ast \, r\cos\theta\,  \psi_m\ .
\end{equation}
We can carry out the integration of $\theta$ analytically.
We perform the numerical integration along $r$-direction and obtain the matrix elements.
Substituting the matrix elements and energy eigenvalues into Eq.(\ref{bnm2}) and using Eq.(\ref{cbb}),
we obtain OTOCs.

\begin{figure}
  \centering
  \subfigure[Microcanonical OTOC]
  {\includegraphics[scale=0.45]{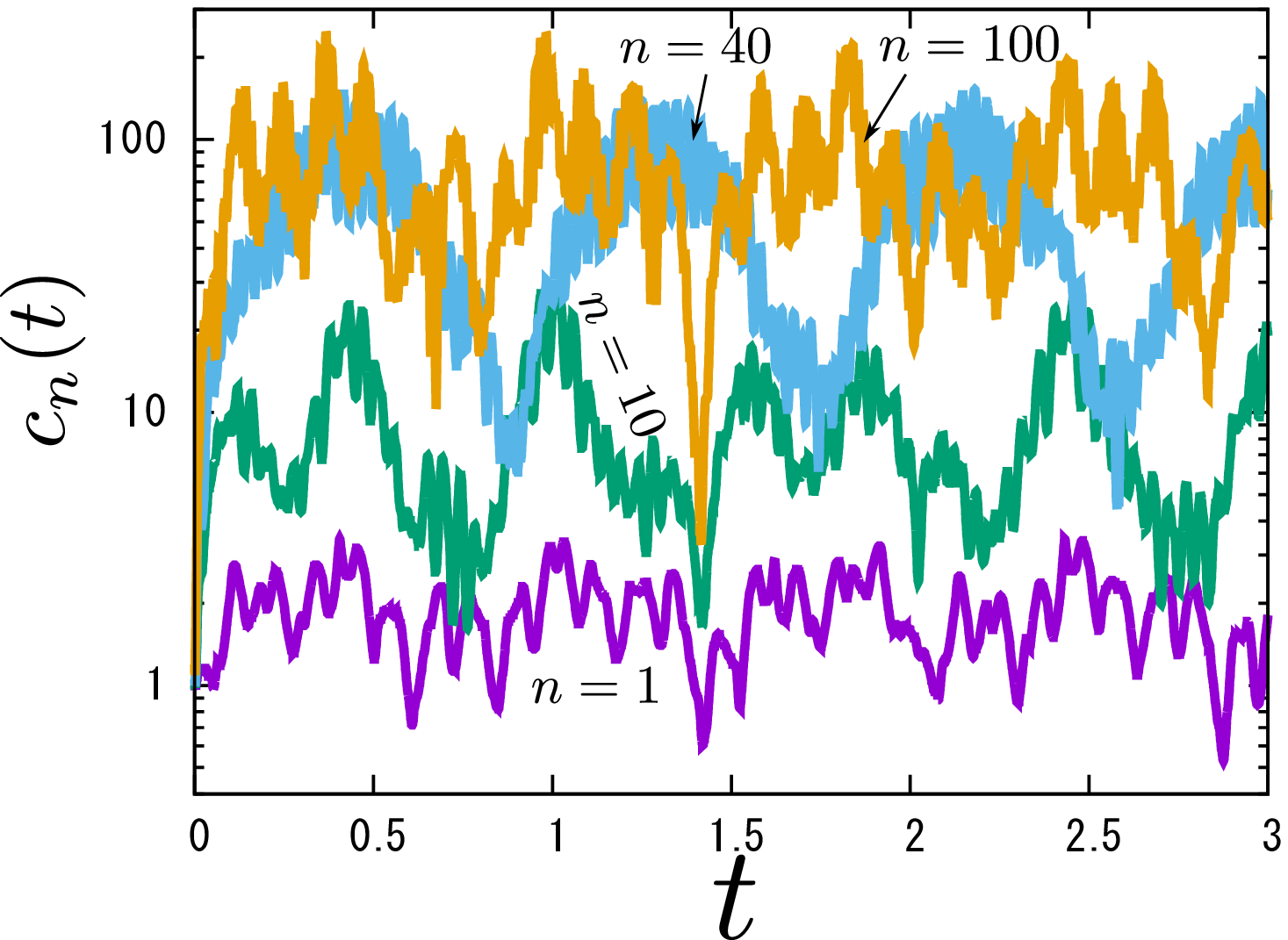}\label{motoccirc}
  }
  \subfigure[Thermal OTOC]
 {\includegraphics[scale=0.45]{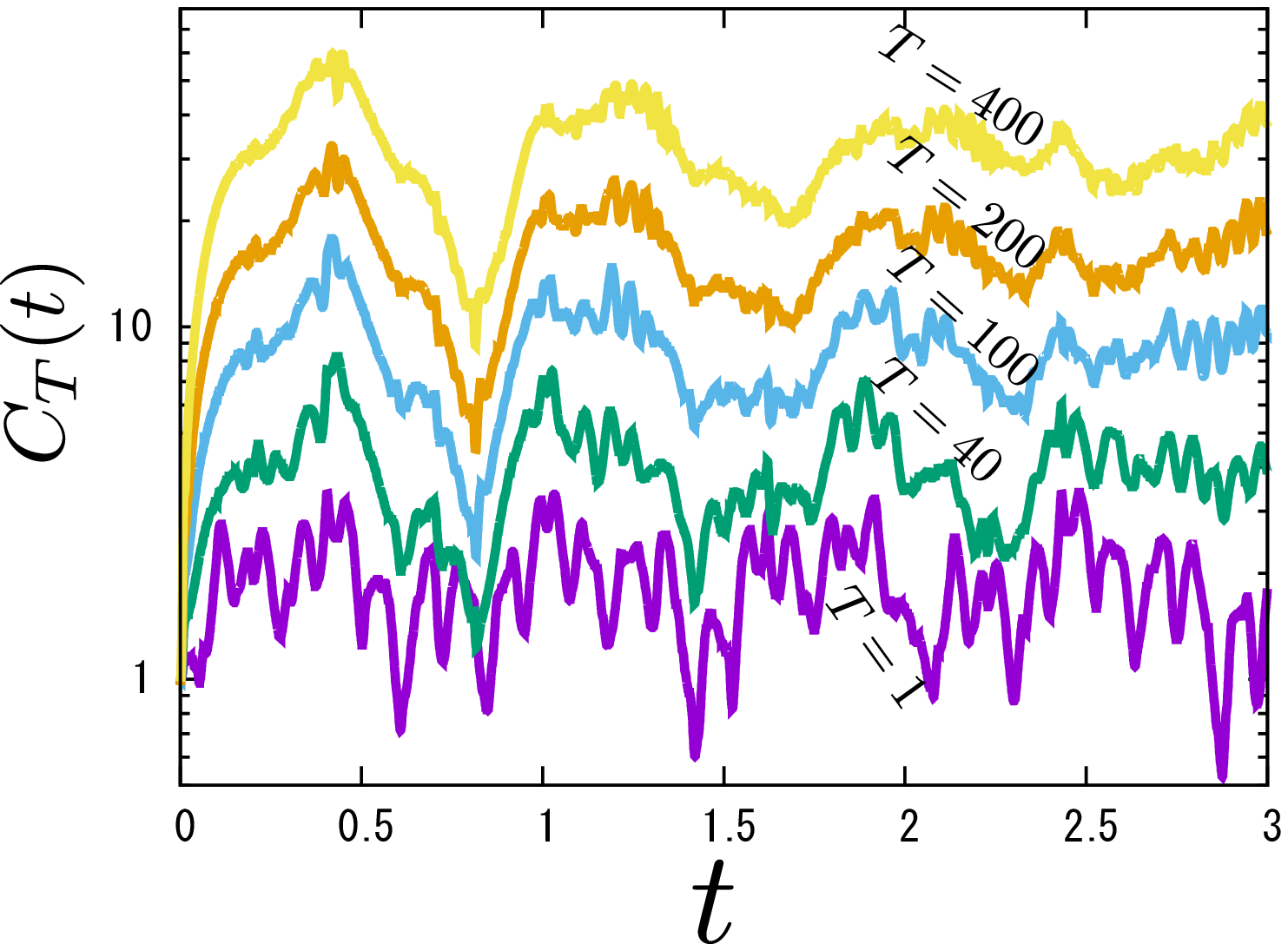}\label{totoccirc}
  }
 \caption{
 Out-of-time-order correlators for the circle billiard.
 }
 \label{otoccirc}
\end{figure}

Fig.\ref{otoccirc} shows the microcanonical and thermal OTOCs for the circle billiard.
The microcanonical OTOCs seem to be non-periodic and tend to be larger for a larger energy level $n$.
We can also find ``dips'' in the microcanonical OTOCs:
For example, for $n=40$ and $100$, they become small ($c_n\sim \mathcal{O}(1)$) around at $t\simeq 0.8$ and $t\simeq 1.4$, respectively.
In the thermal OTOCs, we can also find similar dips around at $t\simeq 0.8$.
We will see that, for the stadium billiard, the dip does not appear in OTOCs.
The dips in OTOCs would originate from the asymptotically commensurable property of the spectrum
and be reflecting the integrability of the systems.

\section{Non-integrable example: Stadium billiards}
\label{sec:non-int}

\subsection{Classical mechanics of stadium billiards}
\label{classical}

\begin{figure}
  \centering
  \subfigure[A typical trajectory]
  {\includegraphics[scale=0.6]{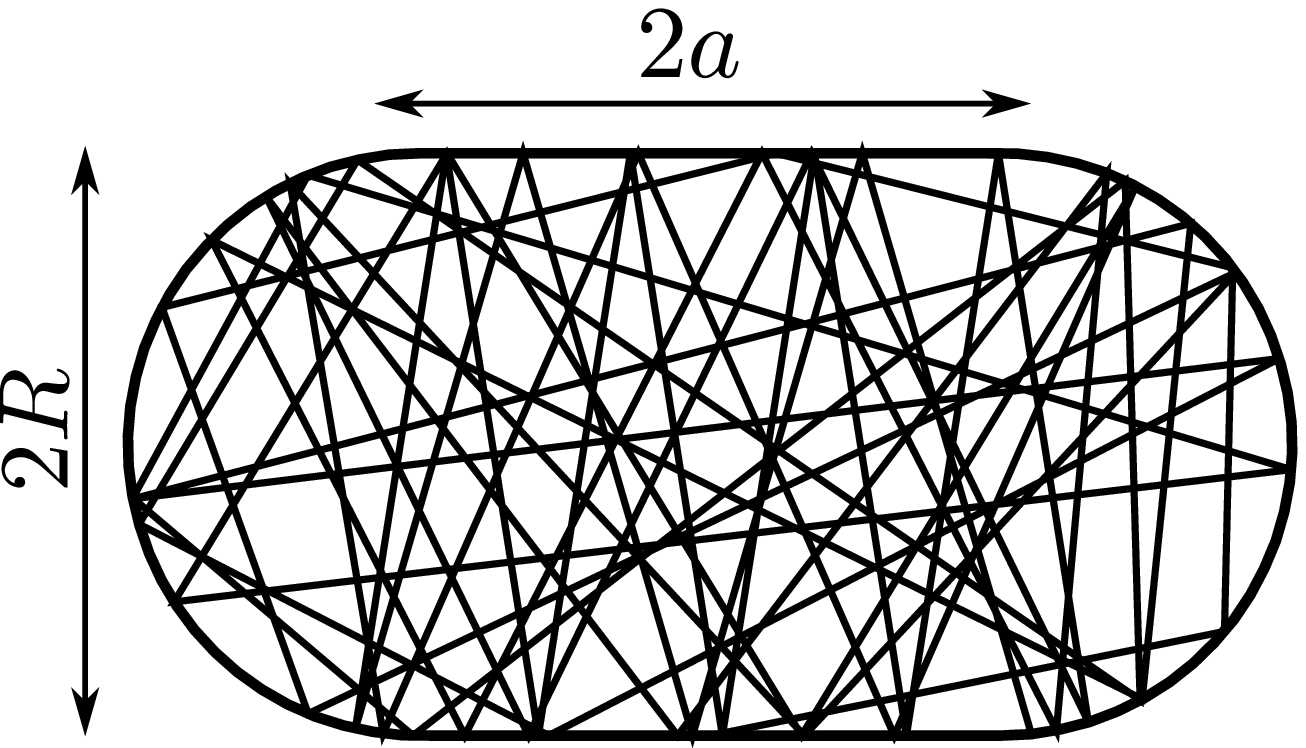}\label{typ}
  }
  \subfigure[Lyapunov exponent]
  {\includegraphics[scale=0.45]{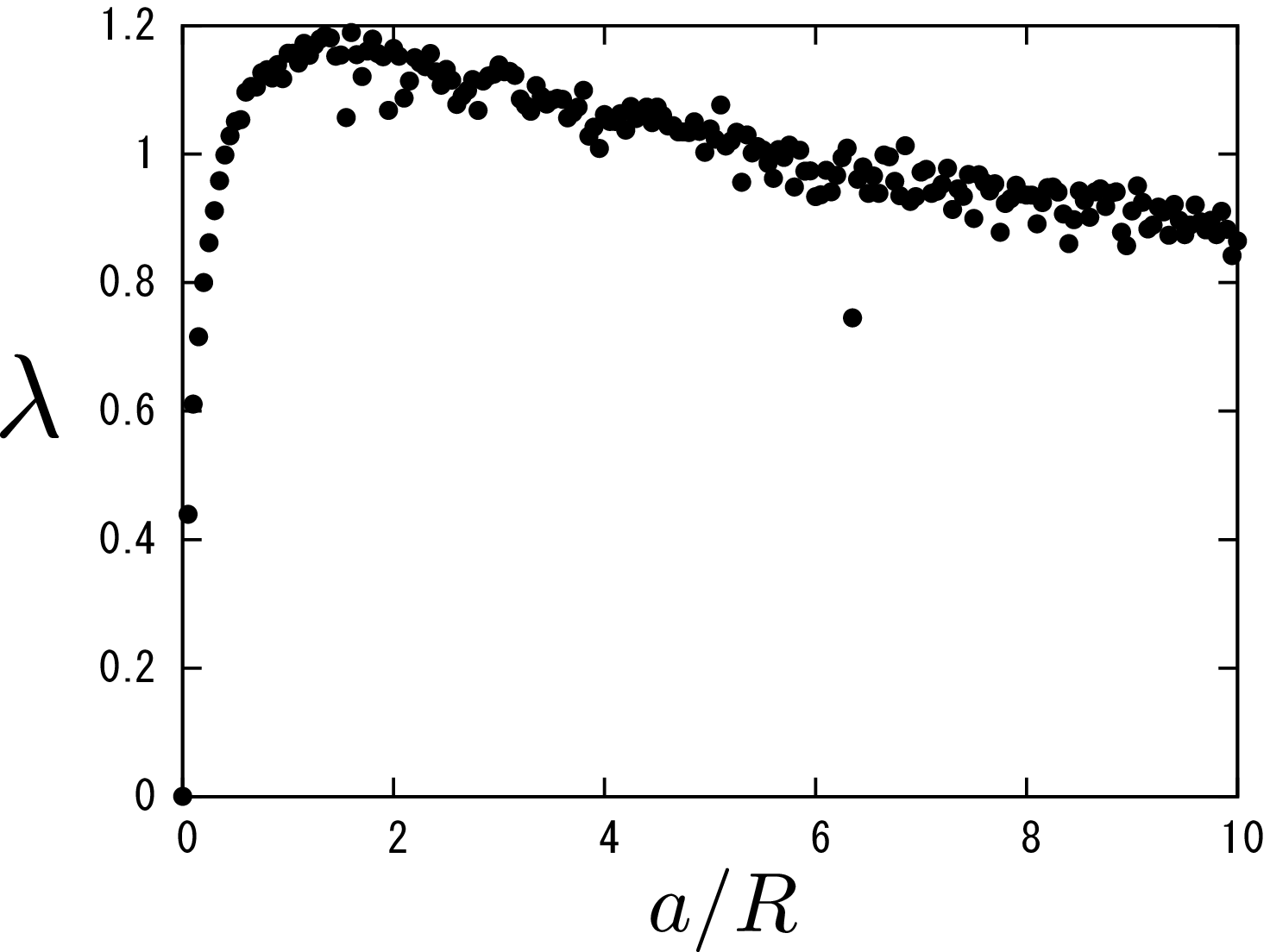}\label{Lyap}
  }
 \caption{
 A typical trajectory of a classical particle in the stadium billiard with $a/R=1$.
 The maximum Lyapunov exponent vs deformation parameter $a/R$
 for fixed area of the stadium and velocity of the particle, $A=v=1$.
 }
 \label{stadium_chaos}
\end{figure}

As a typical example of the non-integrable (chaotic) system,
we consider a stadium billiard~\cite{Sinai,Bunimovich1,Bunimovich2,Bunimovich3,Benettin}:
\begin{equation}
 H=p_1^2+p_2^2+V_\textrm{stad}(x,y) ,\qquad V_\textrm{stad}(x,y)=
 \begin{cases}
  0 & (x,y)\in \Omega\\
  \infty & \textrm{else}
 \end{cases}\ .
\end{equation}
The domain $\Omega$ is shown in Fig.\ref{typ}. 
We denote radii of semicircles as $R$ and the length of straight lines as $2a$.
Let us revisit the classical dynamics of the particle in the billiard. 
Inside the stadium, the particle moves freely with a constant velocity.
At the boundary of the stadium, the particle is reflected elastically.
Fig.\ref{typ} also shows a typical trajectory of the particle in the stadium.
We can find the chaotic behavior.

One of the most characteristic behavior in chaotic systems is the sensitivity to initial conditions:
A tiny deviation of the initial condition causes a significant difference in the future.
The Lyapunov exponent is a useful quantity to measure the strength of the sensitivity to initial conditions. 
Denoting the phase space variable as $\bm{X}(t)$, we consider its linear perturbation:
$\bm{X}(t)\to \bm{X}(t)+\bm{\delta}(t)$.
If $\bm{X}(t)$ is a chaotic solution, because of the sensitivity to initial conditions,
the perturbation expands exponentially as
$\bm{\delta}(t)\sim e^{\lambda t}$. The growing rate $\lambda$ is called Lyapunov exponent.
A positive Lyapunov exponent is the signal of chaos.

In Fig.\ref{Lyap}, we show the Lyapunov exponent as a function of the deformation parameter $a/R$.\footnote{
The boundary condition for the perturbation $\bm{\delta}(t)$ at elastic hard collisions has been studied in Ref.\cite{Dellago}.
We computed the time evolution of $\bm{\delta}(t)$ using the boundary condition. 
}
Here, we took the unit of $v=A=1$,
where $v$ is the velocity of the particle and $A=\pi R^2 + 4aR$ is the area of the stadium.
From the dimensional analysis, we can easily restore $v$ and $A$ by replacing $\lambda\to \sqrt{A} \lambda/v$.
The Lyapunov exponent is zero at the circle limit $a/R=0$. Hence, the classical circle billiard is integrable.
For positive $a/R$, $\lambda$ increases quickly and reaches maximum value around at $a/R\sim1.3$.
The rough estimation of the Lyapunov exponent is 
\begin{equation}
 \lambda \sim \frac{v}{\sqrt{A}}\ ,\quad (a/R\sim 1)\ ,
  \label{LyapRough}
\end{equation}
where $v$ and $A$ are restored. In case of the dynamical billiard,
the Lyapunov exponent is proportional to the velocity $v$, apparently.
(The frequency  of collisions is proportional to the velocity.)
In the squeezed limit $a/R\to \infty$, the particle does not have any chance to hit the semicircles of the stadium.
Thus, $\lambda$ also approaches zero in this limit.
This result is consistent with earlier calculations of Lyapunov exponents in Refs.\cite{Benettin,Dellago,Biham}.

\subsection{Quantum mechanics of stadium billiards}
\label{qbill}

\begin{figure}
\begin{center}
\includegraphics[scale=0.45]{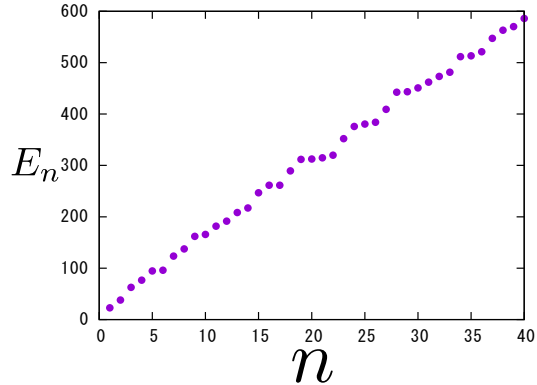}
\end{center}
\caption{
Eigenvalues of the quantum stadium billiard with $a/R=1$.
}
\label{ev_stdium}
\end{figure}

\begin{figure}
  \centering
  \subfigure[$n=1$: $E=2.27\times 10^1$]
  {\includegraphics[scale=0.46]{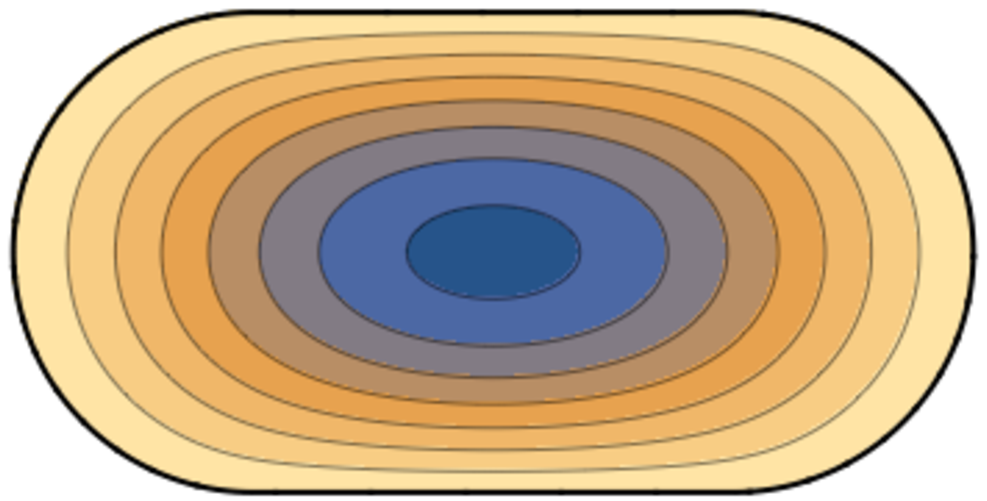}\label{n1}
 }
 \subfigure[$n=2$: $E=3.80\times 10^1$]
  {\includegraphics[scale=0.46]{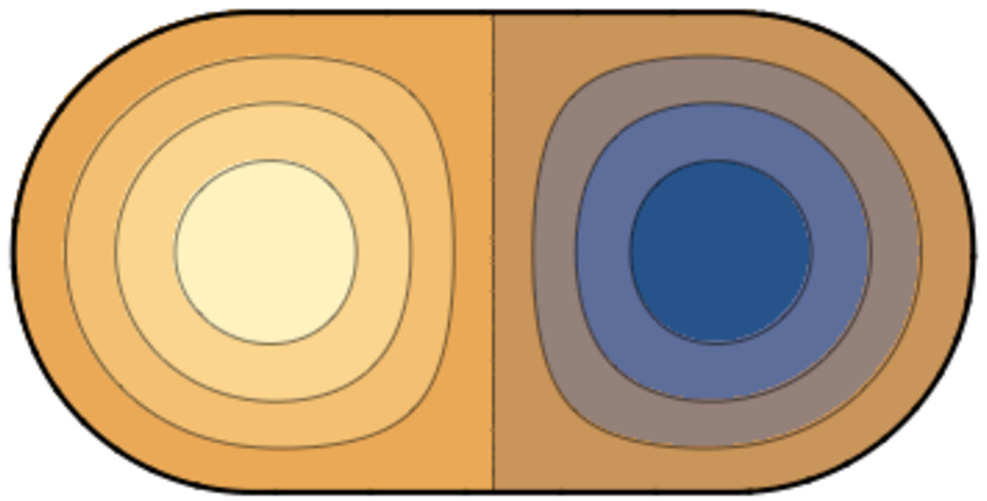}\label{n2}
 }
 \subfigure[$n=7$: $E=1.24\times 10^2$]
  {\includegraphics[scale=0.46]{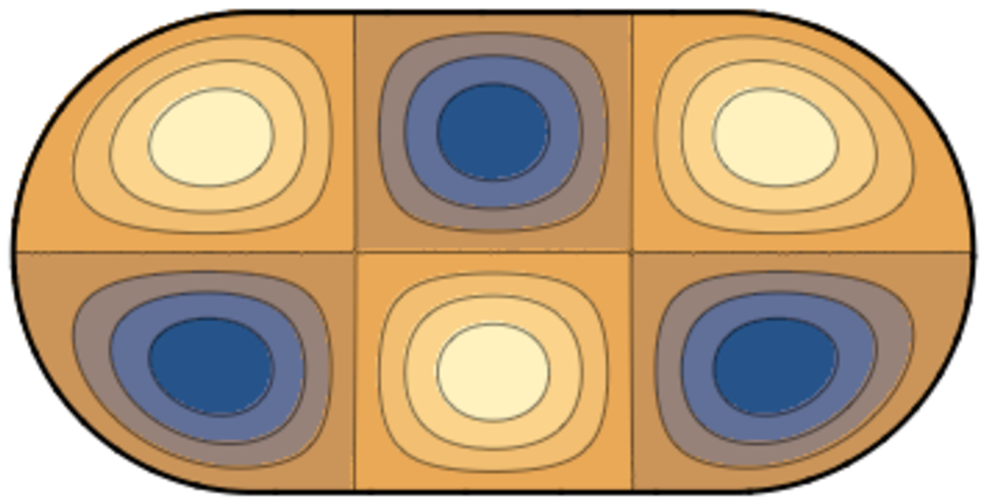}\label{n7}
 }
 \subfigure[$n=50$: $E=7.28\times 10^2$]
  {\includegraphics[scale=0.46]{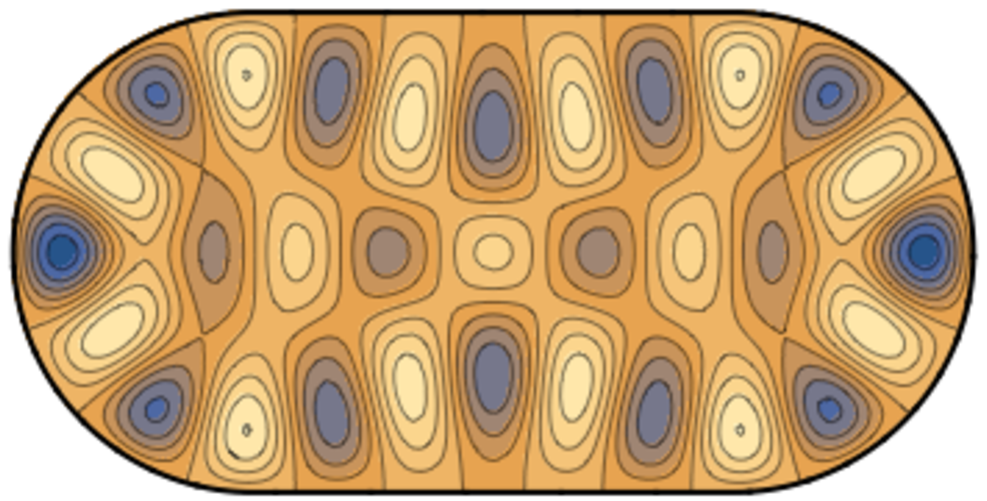}\label{n50}
 }
 \subfigure[$n=200$: $E=2.72\times 10^3$]
  {\includegraphics[scale=0.46]{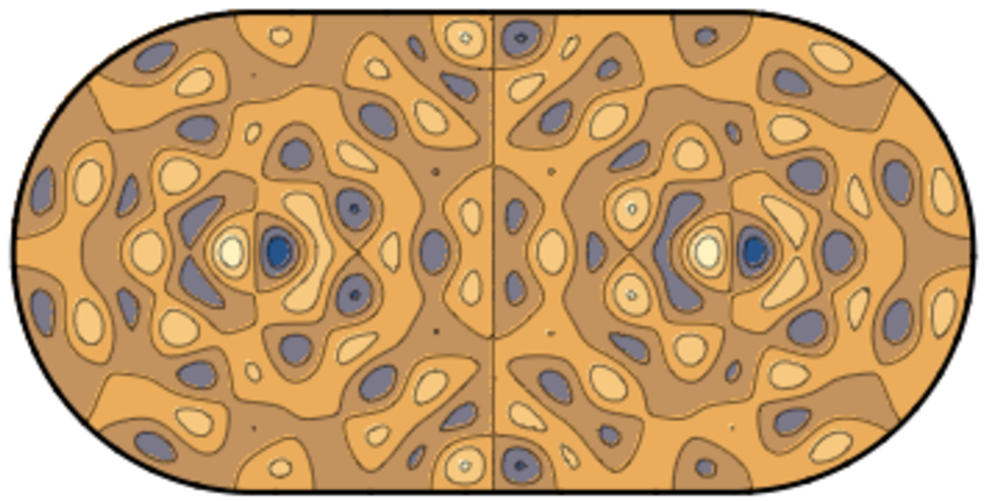}\label{n200}
 }
 \subfigure[$n=400$: $E=5.29\times 10^3$]
  {\includegraphics[scale=0.46]{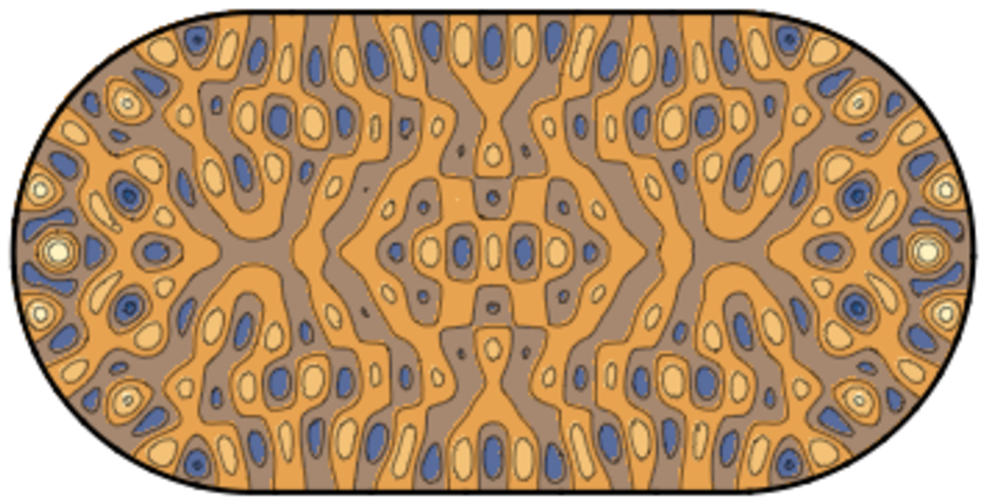}\label{n400}
  }
 \caption{
 Eigenfunctions of the quantum billiard with $a/R=1$ for $n=1,2,7,50,200,400$.
 Corresponding eigenvalues are shown below each figures.
 }
 \label{fig:ef}
\end{figure}

As the quantum version of the dynamical billiard,
we consider the time-independent Schr\"{o}dinger equation 
$[-\Delta+V_\textrm{stad}(x,y)]\psi_n=E_n \psi_n$~\cite{McDonald}.
To determine the eigenvalues and eigenfunctions, 
we used the Mathematica standard package, NDEigensystem, 
which solves eigenvalue problems of linear differential operators using the finite element method.
In Fig.\ref{ev_stdium}, we plot eigenvalues of the quantum stadium billiard with $a/R=1$. 
The energy spectrum is roughly linear in the energy level $n$. 
Fitting the spectrum, we have $E_n\simeq 13 n$ ($a/R=1$). This approximate formula
is useful for rough estimation of the energy spectrum.
In Fig.\ref{fig:ef}, we show eigenfunctions of the quantum billiard with $a/R=1$.

It is known that the Eherenfest time, at which the wave function spreads over the whole system,
becomes quite small for the chaotic system~\cite{Berry,Zurek1,Zurek2,Zurek3}.
To illustrate it, we consider the macroscopic billiard: 
$m=1$ kg, $A=1$ m${}^2$ and $v=1$ m/s. For these parameters, from Eq.(\ref{LyapRough}),
the Lyapunov exponent is estimated as $\lambda\sim 1$ Hz.
Such billiard seems sufficiently classical, but actually, there is tiny uncertainty in its position and momentum.
Let us take the uncertainty as $\Delta x \sim 10^{-17}$ m and $\Delta p \sim 10^{-17}$ kg m/s
so that the uncertainty principle is saturated: $\Delta x \Delta p\sim \hbar$.
In the chaotic system, the wave packet of the particle would exponentially spread as $\ell(t)\sim \Delta x e^{\lambda t}$.
When the size of the wave packet becomes the same order as the system size, $L\sim 1$ m,
a quantum interference effect becomes significant.
The Eherenfest time is estimated as 
$t_E\sim \lambda^{-1}\ln (L/\Delta x)\sim 40$ s.
So even if we start from the extremely localized wave packet, just after one minute,
the system becomes completely quantum.
This behavior is different from what we find in nature.
The problem was that we assumed that the system is isolated from the environment.
Once we take into account the weak interaction between the system and environment,
decoherence is caused and ``the decoherence suppresses the quantum suppression of the chaos''~\cite{Berry}.
For instance, the emergence of the classical chaos due to the decoherence is discussed by considering the continuous quantum measurement \cite{Bhattacharya}.

In this paper, we consider the isolated quantum billiard.
Even for high temperature or high energy, after the Eherenfest time,
the quantum effects will be important and classical approximation will breakdown.
We will revisit this point in section~\ref{sec:Diss}.

\subsection{Out-of-time-order correlators}

\begin{figure}
\begin{center}
\includegraphics[scale=0.5]{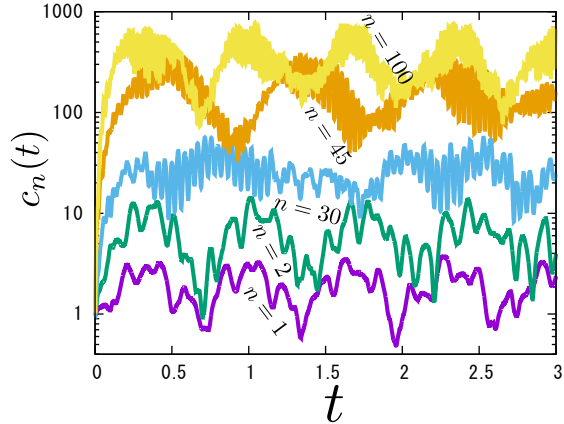}
\end{center}
 \caption{
 Microcanonical OTOC for the stadium billiard with $a/R=1$.
}
\label{motoc_long}
\end{figure}

From the eigenfunctions, we obtain matrix elements of $x$ as
$x_{nm}=\int_\Omega dxdy\, \psi_n x \psi_m$.
Substituting $x_{nm}$ and $E_n$ into Eq.(\ref{bnm2}) and using Eq.(\ref{cbb}),
we compute the microcanonical OTOCs as functions of $t$
for each energy level $n$.
In Fig.\ref{motoc_long}, we show the microcanonical OTOCs for the stadium billiard with $a/R=1$.
For $n=1,2$, OTOCs look similar to those for the particle in a box. (See Fig.\ref{motocbox}.)
This is because typical scales of the wave functions for small $n$ are of the same size as that of the system.
So, wave functions do not ``feel'' the curvature of semicircles of the stadium. 
For higher $n$, however, OTOCs become less recursive than that for the circle billiard
and oscillate around constant values at late time. 
Taking the thermal average of the microcanonical OTOCs, we compute the thermal OTOC.
In Fig,\ref{th_otoc_long}, we show the thermal OTOC for the stadium billiard.
For low temperatures, the lower $n$ mode dominates the thermal OTOC and it looks similar to the microcanonical OTOC for $n=1$.
For high temperature, the thermal OTOC increases quickly as a function of $t$
and approaches a constant value at late time. 
The magnitude of the oscillation around the constant value is small compared to the OTOC of the circle billiard.
In particular, we do not observe dips found in the circle billiard.
We have done same calculations for $a/R=0.2i$ ($i=1,2,\cdots, 10$)
and found qualitatively similar behavior.

\begin{figure}
  \centering
  \subfigure[Stadium billiard ($a/R=1$)]
  {\includegraphics[scale=0.45]{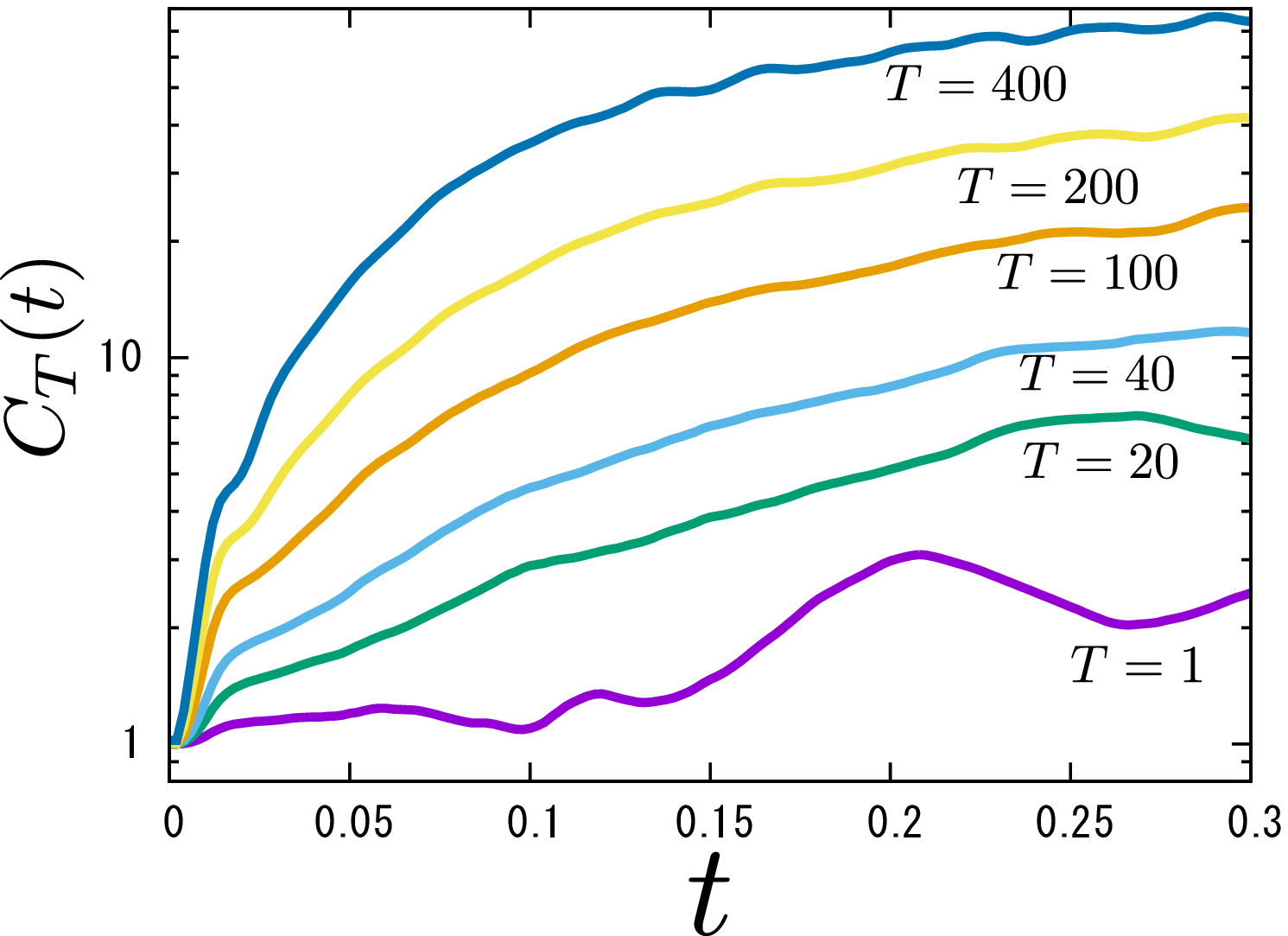}\label{a1_short}
  }
  \subfigure[Circle billiard ($a/R=0$)]
 {\includegraphics[scale=0.45]{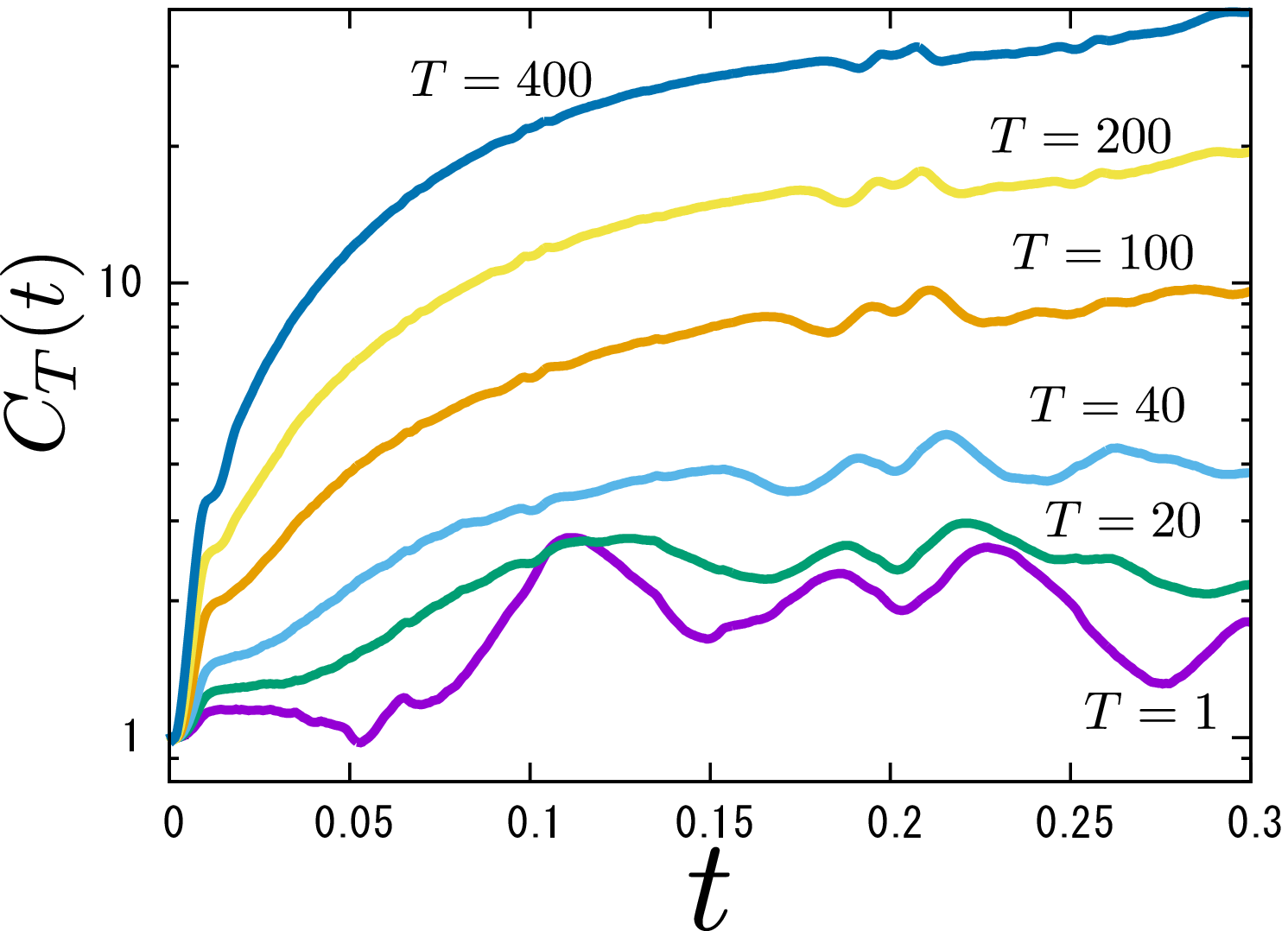}\label{a0_short}
  }
 \caption{
 Early time evolution of thermal OTOCs for the stadium ($a/R=1$) and circle ($a/R=0$) billiard.
 }
 \label{otoc_stdium_short}
\end{figure}

Can we find an exponential growth in the OTOCs?
Fig.\ref{otoc_stdium_short} shows an early time evolution of thermal OTOCs for stadium and circle billiards.
The OTOC for the stadium billiard does not show a clear exponential growth.
At a very early time $t\lesssim 0.01$, one may be able to argue that there is an exponential region.
However, to find the exponential growth $e^{\lambda t}$, we need much longer time than $1/\lambda$.
(Otherwise, we cannot distinguish the exponential and linear functions.)
Moreover, a similarly-looking behavior can be found even for the circle billiard.
There is no qualitative difference in early time OTOCs between the stadium and the circle billiards. 
Our results indicate that, at least for $T\lesssim 400$,
we cannot distinguish integrable and chaotic systems from the early time evolution of the thermal OTOCs.
In Ref.\cite{Maldacena:2015waa}, it was proposed that the Lyapunov exponent $\lambda$ defined by
$C_T(t) \sim e^{2\lambda t}$ satisfies a bound $\lambda\leq 2\pi T$.
The thermal OTOC of the stadium billiard does not show the exponential growth and, in that sense, 
it trivially satisfies the bound.

\begin{figure}
  \centering
  \subfigure
  {\includegraphics[scale=0.45]{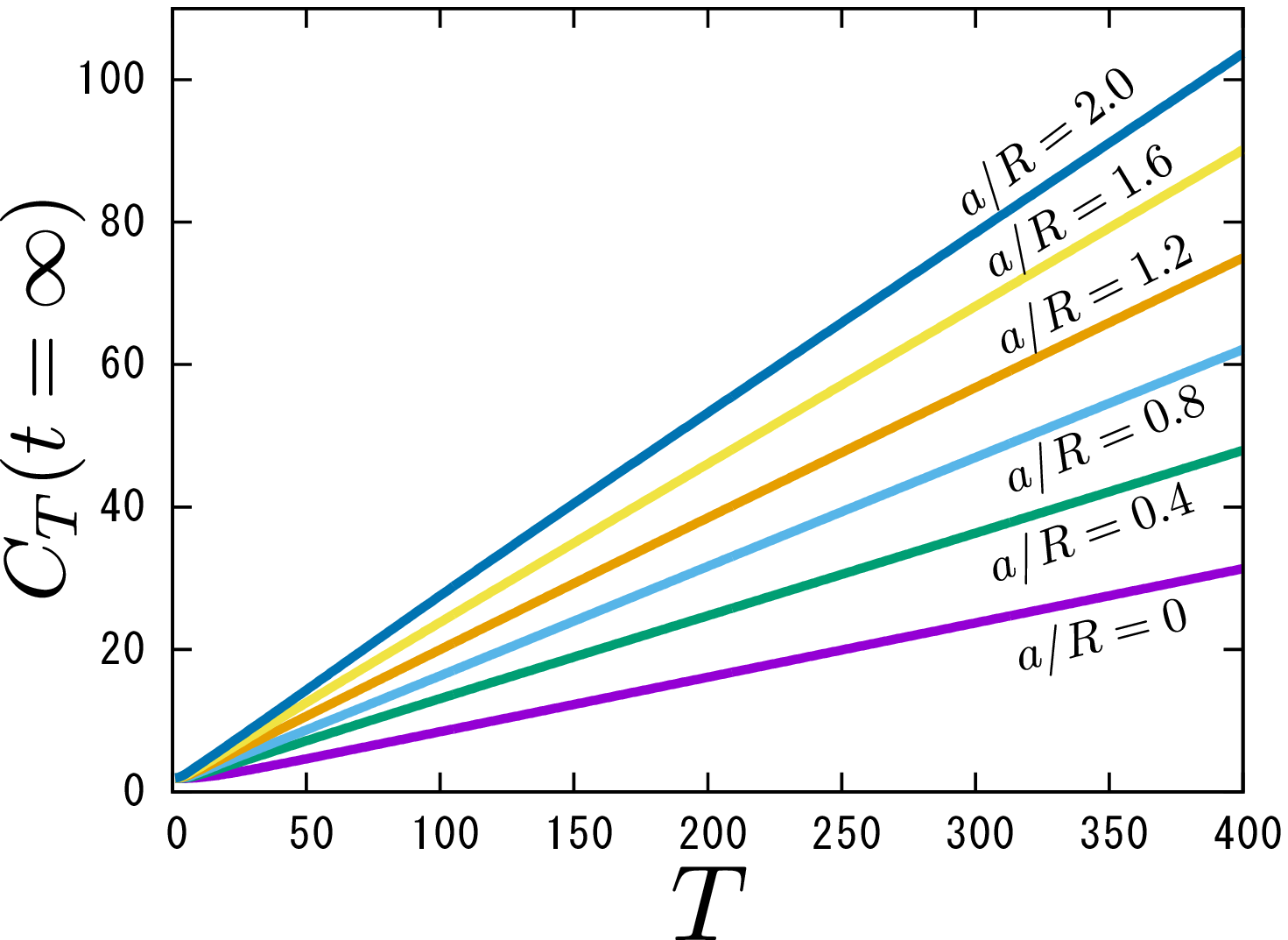}\label{FigCTinf}
  }
  \subfigure
 {\includegraphics[scale=0.45]{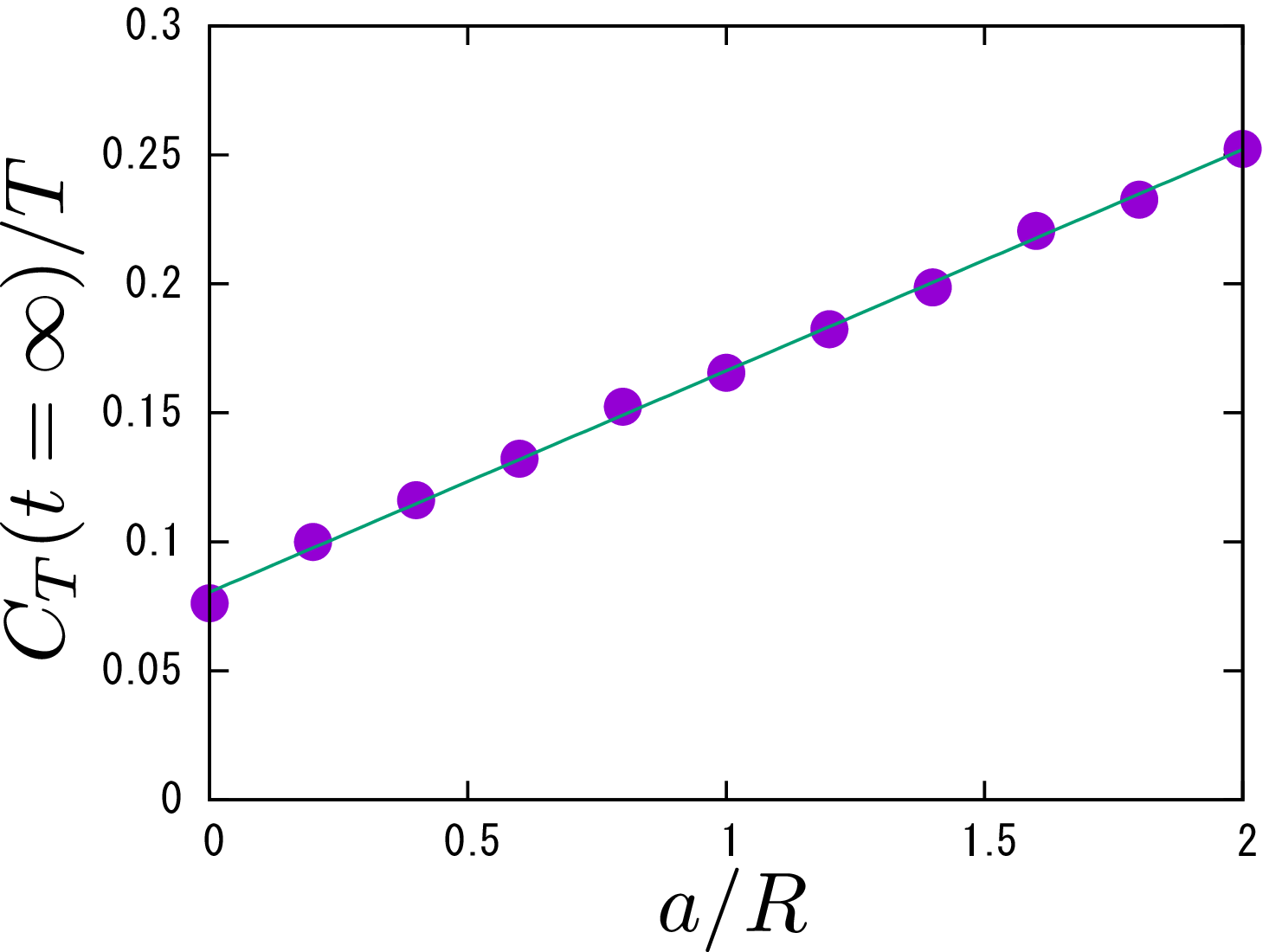}\label{slope}
  }
 \caption{
 (Left) Asymptotic values of thermal OTOC as functions of $T$ for several $a/R$.
 (Right) Plot of their slopes as function of $a/R$.
 }
 \label{otoc_asym}
\end{figure}

We can observe that thermal OTOCs approach constant values at late times.
What determines the asymptotic value?
A naive expectation is that the OTOC saturates when it becomes the ``system size''.
Since the OTOC has the dimension of $\hbar^2$, 
the asymptotic value of the OTOC would be given by $C_T\sim P_\textrm{sys}^2 L_\textrm{sys}^2$ where
$P_\textrm{sys}$ and $L_\textrm{sys}$ are the typical momentum and size of the system.
For a thermal system with the temperature $T$, the typical momentum would be $P_\textrm{sys}\sim \sqrt{mT}$ where
$m$ is the mass of the particle.
Therefore, our expectation is
\begin{equation}
 C_T(t=\infty) \sim m T L_\textrm{sys}^2\ .
  \label{Cinf_conj}
\end{equation}
We can numerically confirm this relation for the stadium billiard
as follows.
We evaluate the asymptotic values of thermal OTOCs from 
$C_T(t=\infty)\simeq \int^{t_2}_{t_2}dt C_T(t)/(t_2-t_1)$. We took $t_1=5$ and $t_2=10$ in actual calculations.
Fig.\ref{FigCTinf} shows $C_T(t=\infty)$ as functions of $T$ for several choices of the 
deformation parameter $a/R$ of the stadium shape.
Our numerical results clearly show that $C_T(t=\infty)$ linearly depends on $T$ and its slope depends on $a/R$.
In Fig.\ref{slope}, we plot the slope $C_T(t=\infty)/T$ as function of $a/R$.
It is also given by a linear function of $a/R$. Fitting the plot, we obtain
\begin{equation}
 C_T(t=\infty)\simeq \left(0.0858\frac{a}{R}+0.0805\right)\times 2mTA\ ,
 \label{CTinf}
\end{equation}
where the area of the billiard $A$ is restored.
Substituting $A=\pi R^2 + 4aR$, we can rewrite above expression as
$C_T(\infty)\simeq 0.68 (a+0.94R)(a+0.79R)mT$.
Since the system size of the stadium is given by $L_\textrm{sys}\sim 2(a+R)$, 
this is consistent with the naive prediction from the dimensional analysis~(\ref{Cinf_conj}).
In the limit of $R\to 0$, the system reduces to the particle in a 1D box with $L=2a$. 
For $R\to 0$, we obtain $C_T(t=\infty)\simeq 0.0858 \times 2m(2a)^2 T$. 
This is also consistent with the time average of the OTOC for the particle in the box~(\ref{Cave_pinb}).

\section{On the classical limit of the out-of-time-order correlator}
\label{sec:Ontheclass}

\subsection{Classical statistics}

First we discuss that the classical statistics does not reproduce the high temperature 
limit of the OTOCs in general.
Using the example of the particle in the 1D box, we can easily show that
classical statistics is not so useful for estimation of the OTOC.
In the high temperature limit,
a naive expectation is that the the thermal average in the OTOC can be replaced by the integral in the 2D phase space as
\begin{equation}
 C_\textrm{cl}(t)=\frac{1}{Z_\textrm{cl}}\int \frac{dx d p}{2\pi} \, e^{-\beta H} \{x(t),p(0)\}_\textrm{PB}^2\ ,
  \label{Ccl}
\end{equation}
where $Z_\textrm{cl}=\int dx d p/(2\pi) \, e^{-\beta H}$ and
$\{\ ,\ \}_\textrm{PB}$ is the Poisson bracket.
For the particle in the box, the classical solution is explicitly written as
\begin{equation}
 x(t)=x(0)+2 p(0) t\ ,\quad p(t)=p(0)\ ,
\end{equation}
before the bounce at a boundary.
After the bounce, the momentum is reflected as $p(t)\to -p(t)$.
We consider the infinitesimal deviation of the initial position fixing the momentum as
$(x(0),p(0))\to (x(0)+\delta x(0),p(0))$.
By the time evolution, the particle will bounce at boundaries.
Then, the deviation of the position change its signature
but the absolute value is constant: $\delta x(t)=(-1)^n \delta x(0)$ after $n$-th bounce.
Therefore, we have
\begin{equation}
 \{x(t),p(0)\}_\textrm{PB}
  =\frac{\delta x(t)}{\delta x(0)}=(-1)^n\ .
\end{equation}
Substituting this into Eq.(\ref{Ccl}), we obtain
\begin{equation}
 C^{\textrm{(box)}}_\textrm{cl}(t)=\frac{1}{Z_\textrm{cl}}\int \frac{dx d p}{2\pi} \, e^{-\beta H} \{(-1)^n\}^2=1\ .
\end{equation}
This classical value is apparently different from
the quantum result of the OTOC at a high temperature
shown in Fig.\ref{otocbox2}.

We can also estimate the classical OTOC for the stadium billiard.
From the sensitivity to initial conditions, we have
$\{x(t),p(0)\}_\textrm{PB}\sim e^{\lambda t}$.
From Eq.(\ref{LyapRough}), the Lyapunov exponent
is $\lambda\sim v\sim p(0)$ for $A=1$.
\begin{equation}
 C_\textrm{cl}^{\textrm{(stad)}}(t)=\frac{1}{Z_\textrm{cl}}\int \frac{d^2x d^2p}{(2\pi)^2} \, e^{-\beta p^2+|p|t}
  =\frac{1}{Z_\textrm{cl}}\int_0^\infty \frac{dp}{2\pi} \,p\,
  e^{-\beta\left(p-\frac{t}{2\beta}\right)^2+\frac{t^2}{4\beta}}
\ . 
  \label{Ccl_stad}
\end{equation}
Although, for fine-tuned initial conditions, the particle motion can integrable,
their measure would be zero.
For $t\gg \beta$, we can replace $\int_0^\infty dp$ by  $\int_{-\infty}^\infty dp$ and we have
\begin{equation}
 C^{\textrm{(stad)}}_\textrm{cl}\sim te^{T t^2}\ .
\end{equation}
This has unusual dependence in $t$ and
is again apparently different from
the quantum calculations in Fig.\ref{th_otoc_long}.

In the case of the harmonic oscillator, on the other hand, 
the classical solutions, $x(t)$ and $p(t)$, are completely identical to Eq.(\ref{xpsol_harm}).
Therefore, classical statistics gives the 
same result as the quantum calculation: $C^{\textrm{(harmonic)}}_\textrm{cl}(t)=\cos^2\omega t$.

\subsection{Out-of-time-order correlator for a wavepacket}
\label{sec:otocw}

Why does quantum statistics not approach classical statistics?
To answer the question, we consider a simpler setup: OTOC for a wavepacket in a 1D box.
We will show that the OTOC deviates from its classical value at a time much earlier than
the Ehrenfest time.

The wavefunction of the wavepacket is given by
\begin{equation}
 \phi(x)=\frac{1}{(2\pi\sigma^2)^{1/4}}\exp\bigg[-\frac{(x-x_0)^2}{4\sigma^2}+ik_0 (x-x_0)\bigg]\ .
  \label{wpacket}
\end{equation}
We consider the well localized wavepacket in real and momentum spaces:
\begin{equation}
 \sigma \ll 1\ ,\qquad k_0 \gg \frac{1}{\sigma}\ .
  \label{local}
\end{equation}
Expanding this wavepacket by the energy eigenstates~(\ref{PBOX_efev}), we obtain 
\begin{equation}
 |\phi\rangle=\sum_n \alpha_n |n\rangle\ ,\qquad
  \alpha_n \simeq i(2\pi\sigma^2)^{1/4}\, e^{-(k_0-\pi n)^2\sigma^2+i(k_0-\pi n)x_0}\ .
  \label{alphan}
\end{equation}
Using the wavepacket, we consider expectation values of commutator $[x(t),p(0)]$ and its square as
\begin{align}
 b_{\phi}&\equiv -i\langle \phi|\,[x(t),p(0)]\,| \phi\rangle=\sum_{n,m}\alpha^\ast_n \,b_{nm}(t)\,\alpha_m\ ,\label{bphi}\\
 c_{\phi}&\equiv -\langle \phi|\,[x(t),p(0)]^2\,| \phi\rangle=\sum_{n,m,k}\alpha^\ast_n\, b_{nk}(t)b_{km}(t)\,\alpha_m\ .\label{cphi}
\end{align}
Here, $b_{nm}(t)$ has been defined in Eqs.(\ref{bnm1}) and (\ref{bnm2}).
We know the analytic expression for the matrix element of $x$~(\ref{xnm_pinb})
and the energy spectrum ~(\ref{PBOX_efev}) for the particle in a box.
We perform the summation numerically.

\begin{figure}
  \centering
  \subfigure[$0\leq k_0 t \leq 0.15$]
  {\includegraphics[scale=0.32]{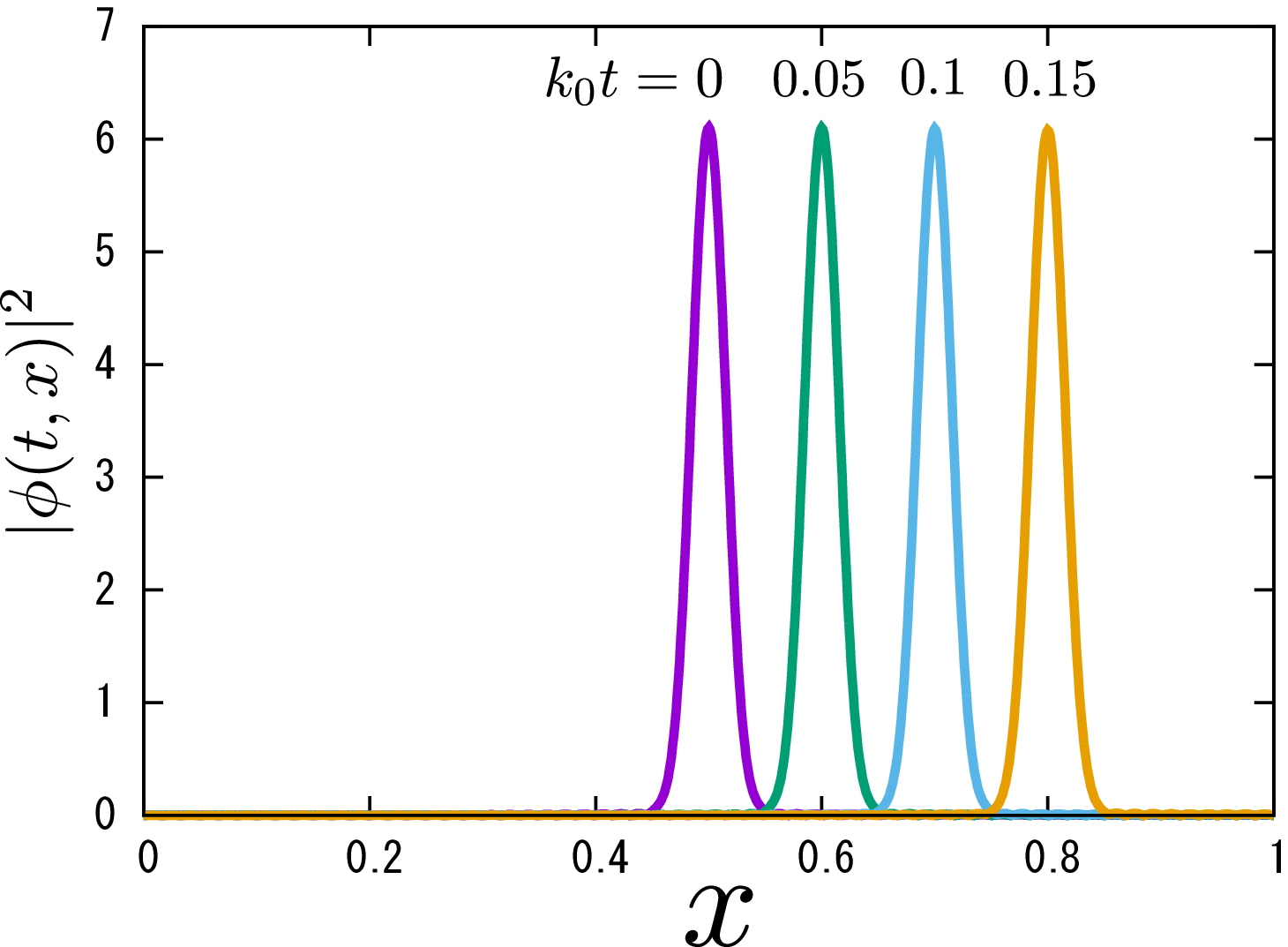}\label{phi_early}
  }
  \subfigure[$5\leq k_0 t \leq 5.15$]
 {\includegraphics[scale=0.32]{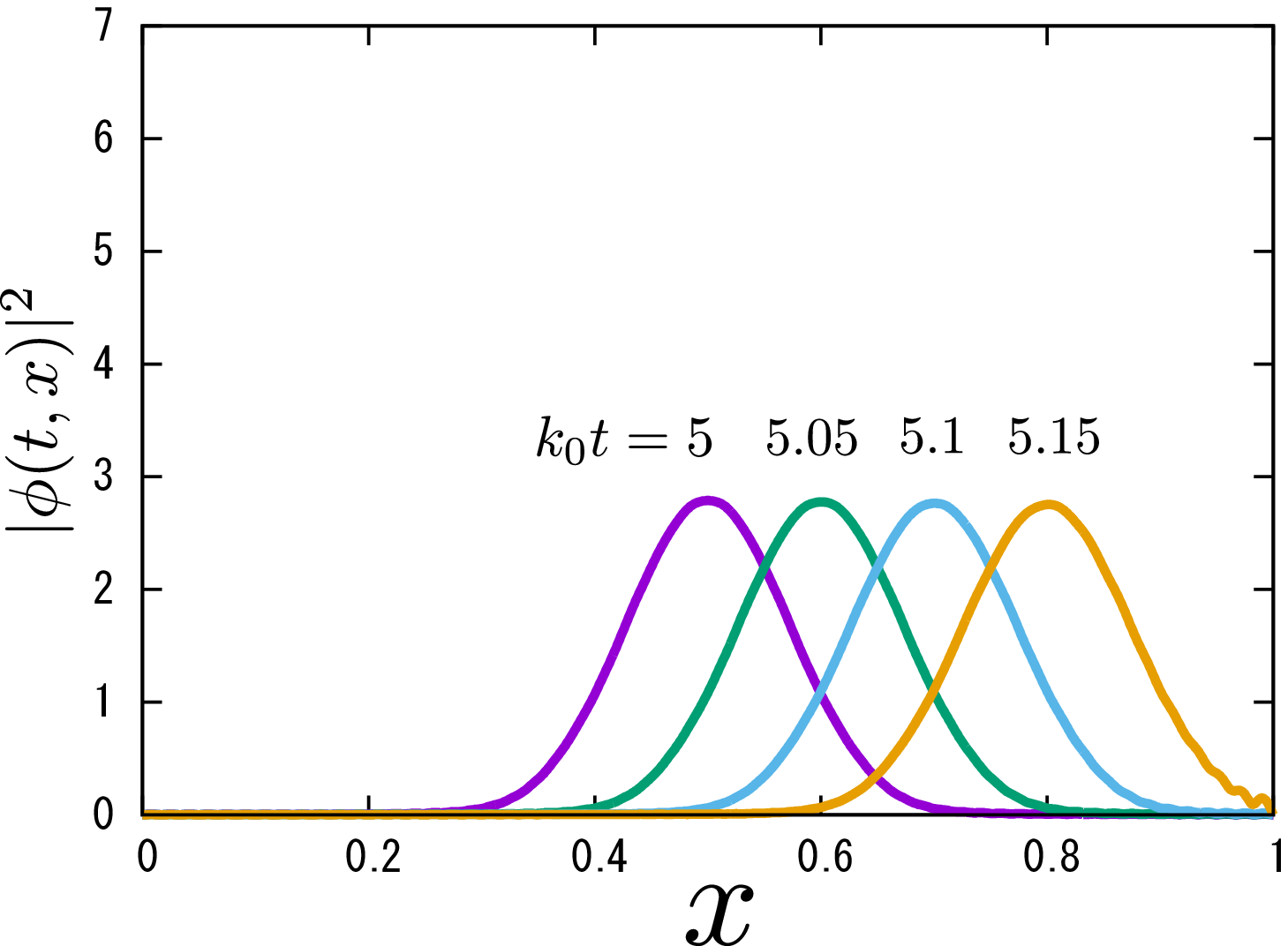}\label{phi_middle}
 }
 \subfigure[$10\leq k_0 t \leq 10.15$]
 {\includegraphics[scale=0.32]{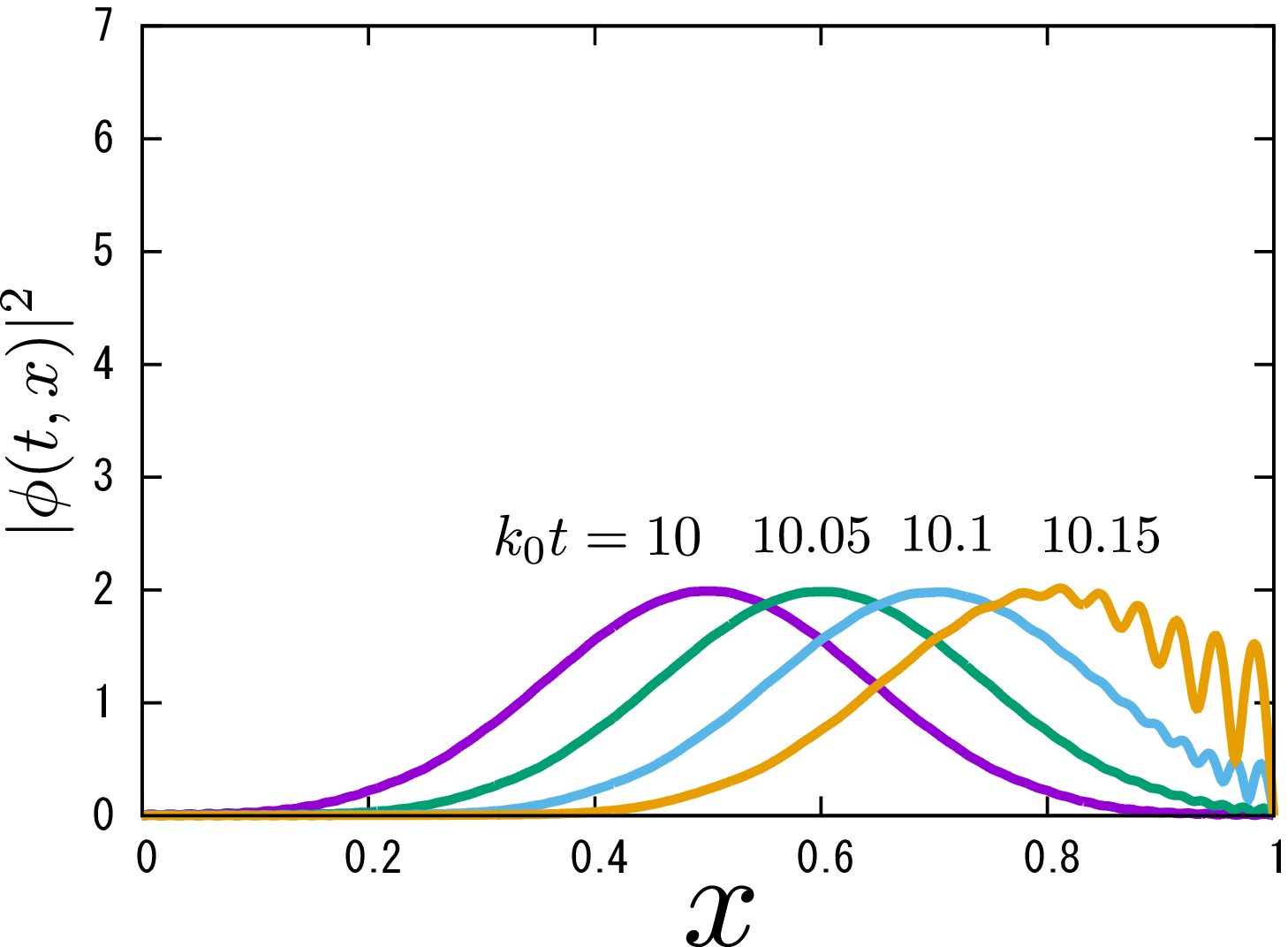}\label{phi_late}
  }
 \caption{
 Time evolution of a wave packet in a box (
 $k_0=3000\pi$ and $\sigma=1/(30\pi)$).
 }
 \label{phi_time}
\end{figure}

We set parameters in the wavepacket as $k_0=3000\pi$, $\sigma=1/(30\pi)$ and $x_0=1/2$.
For $|n-3000|\leq 120$,
we use the expression in Eq.(\ref{alphan}) as $\alpha_n$.
For $|n-3000|> 120$, we simply set $\alpha_n=0$.
Fig.\ref{phi_time} shows the time evolution of the wavepacket $|\phi(t,x)|^2$.
At the early time, the wavepacket is well localized in the real space and shifts
with a constant velocity $v=k_0/m=2k_0$. The wavepacket is getting spread as time increases.
The width of the wavepacket $\sigma_t$ spreads as 
\begin{equation}
 \sigma_t^2=\sigma^2+\frac{t^2}{\sigma^2}\ .
  \label{sigmat}
\end{equation}
(See Eq.(\ref{phifreet}) in the appendix.)
When the width of the wavepacket is the same order as the system size,
a quantum interference effect becomes important. 
Hence the classical (particle) interpretation is no longer valid. 
This time scale is called Ehrenfest time.
For the particle in a box, the Ehrenfest time $t_E$ is estimated from $\sigma_t|_{t=t_E} \sim (\textrm{system size})\sim 1$
and we have
\begin{equation}
 t_E\sim \sigma\ .
  \label{tE_pinb}
\end{equation}

\begin{figure}
  \centering
  \subfigure
  {\includegraphics[scale=0.45]{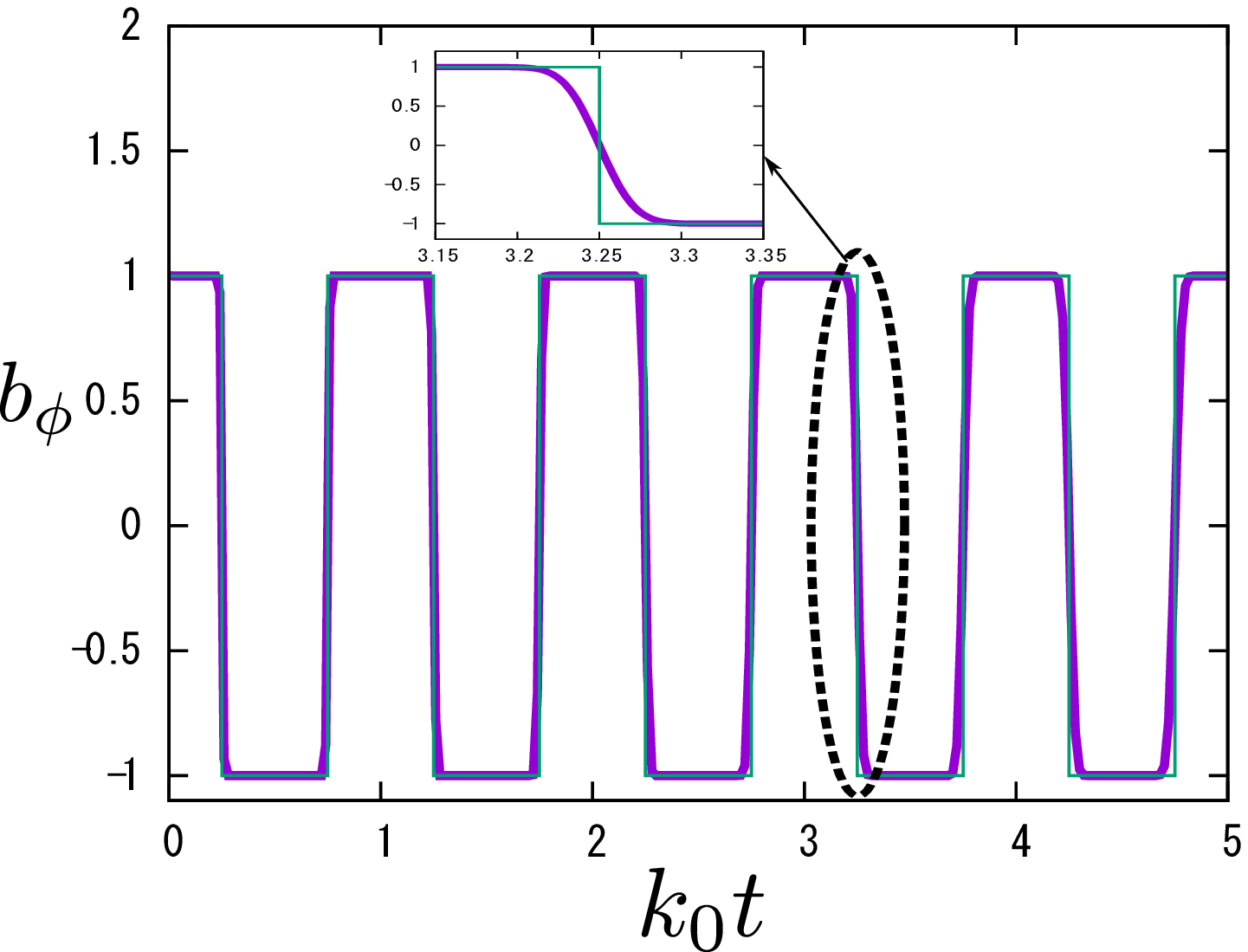}\label{bphifig}
  }
  \subfigure
 {\includegraphics[scale=0.45]{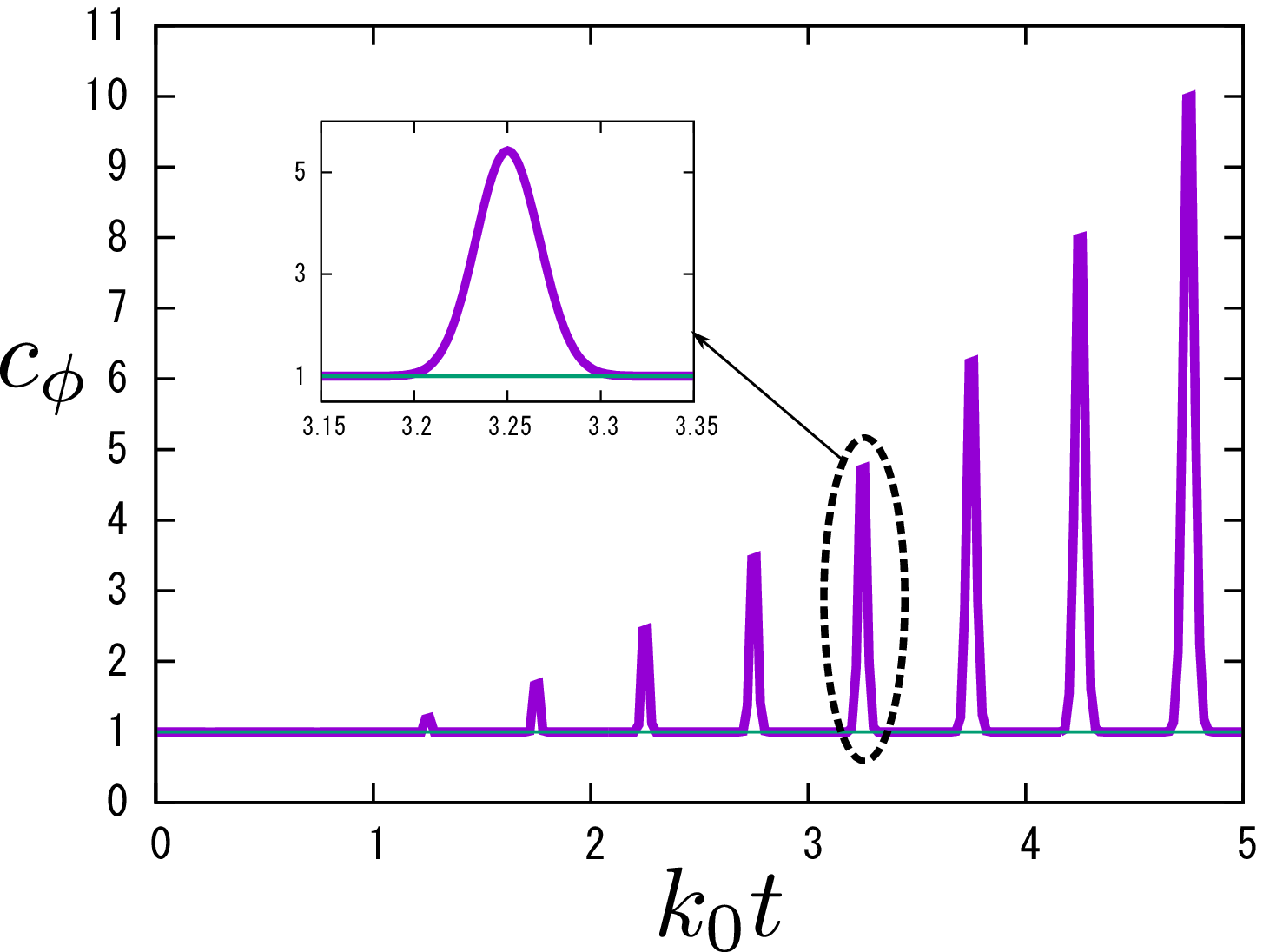}\label{cphifig}
 }
 \caption{
 Time dependence of $b_\phi$ and $c_\phi$.
 Classical predictions for $b_\phi$ and $c_\phi$ are shown by green lines.
 }
 \label{bcphi}
\end{figure}

We evaluate 2- and 4-point OTOCs, $b_\phi$ and $c_\phi$, using the wavepacket.
In Eqs.(\ref{bnm2}) and (\ref{cphi}), as the domain of summation of $k$,
we took $|k-3000|\leq 1000$.
In Fig.\ref{bcphi}, we show the time dependence of $b_\phi$ and $c_\phi$.
In the figure,
classical predictions for $b_\phi$ and $c_\phi$ are shown by green lines:
$b_\phi^\textrm{classical}=(-1)^n$ and $c_\phi^\textrm{classical}=1$.
For the 2-point OTOC $b_\phi$,
the quantum computation from Eq.(\ref{bphi}) nicely coincides with the classical prediction.
In fact, by an analytic calculation in appendix.\ref{app:OTOCw}, we obtain 
\begin{equation}
 b_\phi=\textrm{erf}\left[\frac{\ell(t)}{\sqrt{2}\sigma_t}\right]\ ,
\end{equation}
around at the $(2N+1)$-th bounce. Here, $\ell(t)=2k_0 t +x_0 -(2N+1)$ is the difference
between the center of the wavepacket and
the right boundary $x=1$.
Before the Ehrenfest time $\sigma_t\ll 1$,
the 2-point OTOC is approximated by a step function.
Therefore,
for a well localized wavepacket in a box,
we have a quantum-classical correspondence:
\begin{equation}
 [x(t),p(0)]\sim i\{x(t),p(0)\}_\textrm{PB}\ ,\quad (t\ll t_E)\ .
\end{equation}

On the other hand, for the 4-point OTOC $c_\phi$, we can observe the spiky profile at the time of the bounce:
$k_0 t =0.25+0.5n$ ($n=1,2,3,\cdots$).
Except for the spiky points, it is well approximated by the classical prediction.
By an analytic calculation in appendix.\ref{app:OTOCw}, for $x_0=1.2$, we obtain
\begin{equation}
 c_\phi \simeq
  1+\left(42.0\, \sigma k_0^2+\frac{6.38}{\sigma^3}\right) t^2 \exp\left[-\frac{\ell(t)^2}{2\sigma_t^2}\right] \ .
\end{equation}
around at the bounce at the boundary.
Spikes in right panel of Fig.\ref{bcphi} are gaussians whose widths are given by
$\sigma_t$. 
We focus on the time just on the bounce: $\ell(t)=0$. 
Then, from the inequality of arithmetic and geometric means, the spike term in above equation becomes
\begin{equation}
 \left(42.0\, \sigma k_0^2+\frac{6.38}{\sigma^3}\right) t^2 \geq 32.7 \frac{k_0}{\sigma} t^2 
\end{equation}
The spike term grows to be the same order as the classical value $c^\textrm{classical}_\phi=1$ by
\begin{equation}
 t_s\sim 0.175 \sqrt{\frac{\sigma}{k_0}}\ .
\end{equation}
This is sufficiently earlier than the Eherenfest time because of $t_s/t_E\sim 1/\sqrt{k_0\sigma}\ll 1$.
Therefore, for the wavepacket in the box, we would be able to say 
\begin{equation}
 [x(t),p(0)]^2\nsim -\{x(t),p(0)\}^2_\textrm{PB}\ ,\quad (t_s \lesssim t\ll t_E)\ .
\end{equation}
We need a shorter time scale $t\ll t_s$ to see the quantum-classical correspondence in the 4-point OTOC.

\section{Discussion}
\label{sec:Diss}

The OTOC of the stadium billiard does not show the exponential growth.
We also found that classical statistics for the OTOC 
does not coincide
with the quantum calculation even for the particle in a 1D box.

For the discussion on the disagreement, 
we have to be careful about the classical limit of OTOCs
because it is expected that 
the classical behavior shows up only at times earlier than the Ehrenfest time $t_E$.
Let us estimate the Ehrenfest time~$t_E$ for the thermal system of the particle in a box.
At a temperature $T$, 
the typical energy of the particle should be $E\sim T$.
So, the typical momentum of the particle is estimated as
$k_0 \sim \sqrt{2mE}\sim \sqrt{T}$. 
Although there is no a priori choice for the typical size of a particle $\sigma$, 
it should satisfy $T^{-1/2}\ll \sigma \ll 1$ from the well-localized-condition~(\ref{local}).
We take the thermal de Broglie length as the typical size of the particle $\sigma\sim T^{-1/2}$ since 
this gives the smallest Ehrenfest time. 
Then, from Eq.(\ref{tE_pinb}),
the Ehrenfest time is estimated as $t_E\sim T^{-1/2}$.
For $T\sim 100$, we have $t_E\simeq 0.1$. 
In Fig.\ref{otocbox2}, even if we focus on the time scale of
$t\ll t_E$, the OTOC disagrees with the classical value $C_\textrm{cl}(t)=1$.

We can argue the stadium billiard in a similar manner.
For the chaotic system with a Lyapunov exponent $\lambda$,
the width of a wavepacket would spread exponentially as
$\sigma_t \sim \sigma e^{\lambda t}$.
The Ehrenfest time, which is estimated from $\sigma_t|_{t=t_E} \sim 1$, is given by
$t_E \sim \lambda^{-1}\ln(\sigma^{-1})$.
For a thermal system, the typical velocity is given by $v=k_0/m\sim \sqrt{T}$. 
Then, from Eq.(\ref{LyapRough}),
the Lyapunov exponent is $\lambda\sim \sqrt{T}$.
Choosing the typical size of the particle as $\sigma\sim T^{-1/2}$ again,
we can estimate the Ehrenfest time as
$t_E \sim T^{-1/2}\ln T$.
For $T=400$, we have $t_E\sim 0.3$.
In our numerical result of the OTOC given 
in Fig.\ref{a1_short}, we cannot find an exponential growth for the time region $t\ll t_E$.

Why do the OTOCs deviate from their classical value at a high temperature and $t\ll t_E$?
In section.\ref{sec:otocw},
we found the other time scale $t_s$,
at which the quantum-classical correspondence of the 4-point OTOC is violated,
for a wavepacket in a box. 
We showed that $t_s$ is sufficiently smaller than the Ehrenfest time $t_E$. 
Although we do not have any physical interpretation of $t_s$ at the moment,
the existence of the time scale $t_s$ would be an origin of the distinction
between quantum and classical mechanics as for the OTOCs. 
The time scale $t_s$ might stem from the interference effect at the bounce (Fig.\ \ref{bcphi}). 
In fact, such a small time scale does not show up in Ref.\cite{Rozenbaum}, 
in which the system without the boundary was considered.

\begin{figure}
\begin{center}
\includegraphics[scale=0.9]{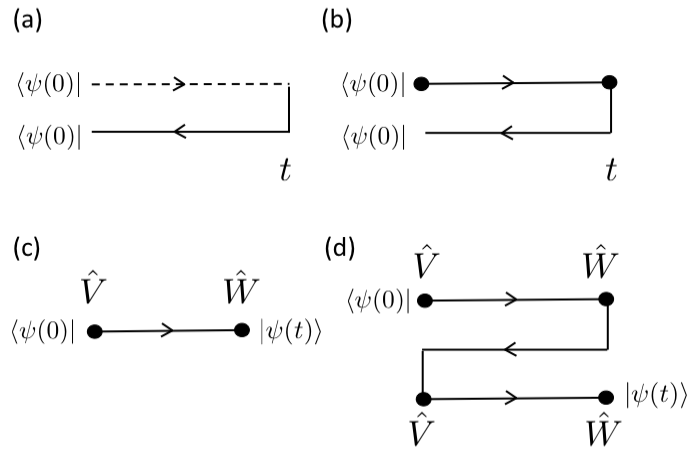}
\end{center}
\caption{
(a) Loschmidt echo. The dashed line is a time evolution to time $t$ by the Hamiltonian 
$H'$, while the solid line is that by the Hamiltonian $H$.
(b) Loschmidt echo, to the second order in perturbation of $H'-H$. The dots represent
the perturbation, which could take place anywhere on the dashed line in (a). We depict 
the case when the perturbations are at both ends, $t=0$ and $t$. (c) A time-order correlator, $\langle \psi | \hat{V} \hat{W}(t) |\psi\rangle = \langle \psi | \hat{V} 
e^{iHt}\hat{W}e^{-iHt} |\psi\rangle$. (d) An OTOC,
$\langle \psi | \hat{V} \hat{W}(t)\hat{V} \hat{W}(t) |\psi\rangle = \langle \psi | 
\hat{V} e^{iHt}\hat{W}e^{-iHt}\hat{V} e^{iHt}\hat{W}e^{-iHt} |\psi\rangle$.
}
\label{Los}
\end{figure}

It has been also 
known for quantum fidelity or Loschmidt echo (see \cite{Prosen} for a review) that generally
the time region for reproducing the classical Lyapunov behavior is quite limited.
The Loschmidt echo 
measures how identical the state is to the state once time-evolved by a Hamiltonian
$H'$ and then time-evolved backward in time by a slightly different Hamiltonian $H$.
As shown in Fig.~\ref{Los}, generic OTOCs 
are interpreted as a generalization of the Loschmidt echo.\footnote{This similarity
was discussed and further generalized in \cite{Aleiner,Campisi,Tsuji,Kurchan,Bohrdt,Tsuji2,Kuku}, for example to a generalized fluctuation-dissipation theorem \cite{Tsuji2}.}
Therefore, it is natural that the semiclassical limit of the OTOC of the billiard
does not reproduce the classical Lyapunov behavior, as in the case of the fidelity.

From asymptotic values of OTOCs for the stadium billiards, 
we found the empirical relation for the typical magnitude of the thermal OTOC:
$C_T \sim m T\times (\textrm{system size})^2$.
This result indicates that
the magnitude of the OTOC does not relate to the magnitude of chaos.
In fact, while the classical Lyapunov exponent has a maximum value around at $a/R=1.3$ as in Fig.\ref{Lyap},
the magnitude of the OTOC is just given by a linear function in $a/R$ for fixed $A$ as in Eq.~(\ref{CTinf}).

By a naive argument in Sec.~\ref{intro}, 
the OTOC can be related to the classical Lyapunov exponent via
the replacement of the commutator by a Poisson bracket.
However, in our analyses of the quantum stadium billiards, we do not find
the exponential growth of the OTOC.
Is there single particle quantum mechanics which shows clear exponential growth in the thermal OTOC?
Can the quantum Lyapunov exponent saturate the bound provided in Ref.\cite{Maldacena:2015waa}?
To answer these questions, 
we need further study of OTOCs of classically chaotic systems.

The OTOC for the Sachdev-Ye-Kitaev (SYK) model grows exponentially~\cite{Sachdev:1992fk,Kitaev-talk-KITP}.
What was essential for the exponential growth?
There are two significant difference between the SYK and our examples.
(1)The OTOC in the SYK model has been calculated in a large $N$ limit while our examples concern single particle quantum mechanics. 
For the large $N$ theory, we can divide the system into two parts, A and B. 
The part B can be regarded as the ``environment'' by integrating out the degree of freedom in B.
The interaction between the environment and part A would cause the decoherence~\cite{Berry,Zurek1,Zurek2,Zurek3}. 
It follows that the system would be classical-like and show the exponential growth in the OTOC. 
Indeed, the emergence of the decoherence by taking the partial trace for the environment have been shown \cite{Zeh, Kiefer}. 
The coupled systems, each of them classically shows the chaotic behavior, has also been investigated and shown to have decoherence effect \cite{Adachi}. 
(2)The SYK model has the random coupling. 
The random coupling is known to enhances the decoherence \cite{Castagnino}. 
Adding to that, it is known that the wave function is localized in space when the system has random potential (Anderson localization). 
The localization of the wave function would be regarded as emergence of particle nature and thus the exponential growth of the OTOC might be expected since the presence of the wave nature in our model prevents us to observe the exponential growth. 
However, the localization is also known to have negative effect for classicalization since the diffusion is suppressed due to the localization in the momentum space \cite{Fishman}. 
Thus the effect of the randomness on OTOC is still unclear.

\acknowledgments

We would like to thank 
Taro Kimura, Michikazu Kobayashi, Yasusada Nambu, Makoto Negoro, and Takahiro Sagawa
for valuable discussion and comments.
The numerical calculations were partially carried out using the TSC-computer of Topological Science in Keio university.
The work of K.H.\ was supported in part by JSPS KAKENHI Grant Numbers 15H03658, 15K13483.
The work of K.M.\ was supported by JSPS KAKENHI Grant Number 15K17658.
The work of R.Y.\ was supported by the MEXT-Supported Program for the Strategic Research Foundation at
Private Universities gTopological Scienceh (Grant No. S1511006).

\appendix

\section{Truncation error}
\label{truncerror}

\begin{figure}
\begin{center}
\includegraphics[scale=0.5]{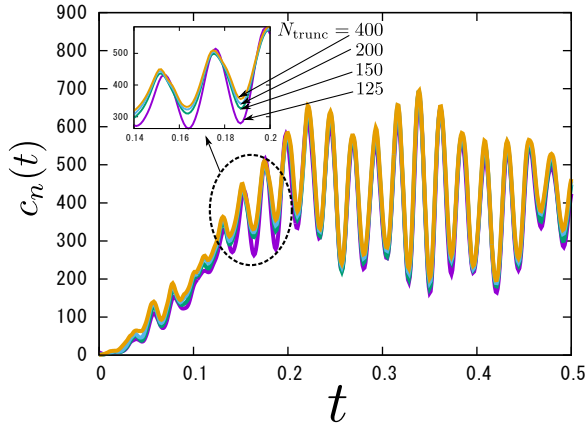}
\end{center}
 \caption{
The microcanonical OTOC of the stadium billiard for $n=100$.
The truncation is varied as $N_\textrm{trunc}=125,150,200,400$.
}
\label{otoc_Ndep}
\end{figure}

In the several places for evaluation of OTOCs~(\ref{bnm2}), (\ref{cbb}) and (\ref{CT}),
we need the summation of infinite terms. 
In the actual numerical calculations, we have truncated the summation at $n=N_\textrm{trunc}$.
In this section, we study the $N_\textrm{trunc}$-dependence of OTOCs.
Here, we focus on the stadium billiard with $a/R=1$.
We consider the microcanonical OTOC with $n=100$.
Fig.\ref{otoc_Ndep} shows the microcanonical OTOC for $N_\textrm{trunc}=125,150,200,400$.
The OTOC nicely converges as $N_\textrm{trunc}$ increases.
For $n<100$, we found better convergence than $n=100$. 
For $n>100$, microcanonical OTOCs does not contribute to the thermal OTOC so much
because of the suppression factor $\exp(-E_n/T)$.
(In this paper, we consider $T\leq 400$ for the stadium billiard.
For $n=100$, the energy eigenvalue is $E_{100}\simeq 1300$ and its contribution is suppressed by $\exp(-E_n/T)\simeq 0.04$.)
Based on the analysis in this section,
we chose $N_\textrm{trunc}=400$ for most of calculations of stadium billiards.

 \section{Analytic calculation of the OTOC for a wavepacket in a box}
 \label{app:OTOCw}

\subsection{Propagation of a wavepacket in a box}

We consider dynamics of a wavepacket in a 1D box: $V(x)=0$ $(0<x<1)$, $\infty$ (else).
The initial gaussian wave function is given by Eq.(\ref{wpacket}). 
We consider the well localized wavepacket satisfying Eq.(\ref{local}).
We also assume that the center of the wavepacket is separated from boundaries:
\begin{equation}
 x_0\gg \sigma\ ,\qquad 1-x_0 \gg \sigma\ .
 \label{x0cond}
\end{equation}
Then, we do not have to mind tiny non-zero values of the wave function at boundaries.
For the free particle $V(x)=0$, time evolution of the wavepacket is given by
\begin{multline}
 \phi_\textrm{free}(t,x)=U_\textrm{free}(t)\phi(x)\\
 =\frac{1}{(2\pi\sigma^2)^{1/4}\alpha(t)^{1/2}}\exp\left[
\frac{1}{\alpha(t)}\left\{-\frac{(x-x_0)^2}{4\sigma^2}+i[k_0 (x-x_0)-k_0^2]\right\}
 \right]
 \label{phifreet}
\end{multline}
where $\alpha(t)=1+it/\sigma^2$ and $U_\textrm{free}(t)$ is the time evolution operator of the free particle:
$U_\textrm{free}(t)=e^{-ip^2t}$. The absolute square of the wavepacket is given by a gaussian:
\begin{equation}
 |\phi_\textrm{free}(t,x)|^2=\frac{1}{(2\pi\sigma_t^2)^{1/2}}\exp
  \left[
   -\frac{\{x-(x_0+2k_0t)\}^2}{2\sigma_t^2}\right]\ .
\end{equation}
where $\sigma_t$ is defined in Eq.(\ref{sigmat}). The width of the wavepacket spreads as a function of $t$.
The center of the wavepacket is moving with a constant velocity $v=2k_0$.

In case of the particle in a box,
the dynamics of the wavepacket is given by the ``folding operation'' of the free wavepacket as 
\begin{equation}
 \phi(t,x)=U(t)\phi(x)
  =
  \sum_{m=-\infty}^\infty
  \left[
\phi_\textrm{free}(t,x+2m)-\phi_\textrm{free}(t,2m-x)
  \right]h(x)\ .
\end{equation}
where $U(t)=\exp[-i(p^2+V(x))t]$ and $h(x)=\theta(x)\theta(1-x)$. 
One can check that this satisfies Schr\"odinger equation $i\partial_t \phi=-\partial_x^2 \phi$
and boundary conditions $\phi(t,x=0,1)=0$. 
Hereafter, we only consider much earlier time than the Ehrenfest time, 
\begin{equation}
 t\ll t_E\sim \sigma\ .
  \label{tE}
\end{equation}

For following calculations, it is convenient to introduce the ``folding operator'' $F$ by
\begin{equation}
 F\chi(x)=\sum_{m=-\infty}^\infty[
\chi(x+2m)-\chi(2m-x)
]h(x)\ .
\end{equation}
Using the folding operator, the time evolution operator for the particle in a box is written as
\begin{equation}
 U(t)=FU_\textrm{free}(t)\ .
\end{equation}
Also the Hermite conjugate of the time evolution operator is written as
$U^\dag(t)=U(-t)=FU_\textrm{free}(-t)=FU^\dag_\textrm{free}(t)$.
One can easily check following formulae of the folding operator:
\begin{align}
&F\chi(x+2n)=F\chi(x)\ ,\label{Fform1}\\
&F\chi(-x)=-F\chi(x)\ , \label{Fform2}
\end{align}
where $n\in \bm{Z}$.

\subsection{Operation of $x(t)$ and $p(0)$ to the wavepacket}

We consider around $(2N+1)$-th bounce of the wavepacket at boundaries:
The center of the free wavepacket $x_\textrm{center}=2k_0 t+x_0$ 
is in $2N<x_\textrm{center} < 2N+2$.
We also assume that the wavepacket does not overlap with the left boundary: $x_\textrm{center}-2N\gg\sigma$ and
$2N+2-x_\textrm{center}\gg\sigma$.
Then, dynamical solution can be approximated by
\begin{equation}
 \phi(t,x)
 \simeq 
\left[\phi_\textrm{free}(t,x_+)-\phi_\textrm{free}(t,x_-)
		\right]h(x)\ ,
\label{Nbounce}
\end{equation}
where we define
\begin{equation}
 x_+=x+2N\ ,\qquad x_-=2N+2-x\ .\label{xpmdef}
\end{equation}

For the computation of the OTOC, we consider the operation of $x(t)$ and $p(0)$ to the wavepacket.
For the initial gaussian wavepacket, we can rewrite the operation of the momentum operator as
\begin{equation}
 p\, \phi(x)=-i\partial_x\phi(x)
  =\mathcal{A}\phi(x)\ ,\qquad \mathcal{A}\equiv \frac{1}{2\sigma^2}\partial_{k_0}+k_0\ .
  \label{Aphi}
\end{equation}
The operator $\mathcal{A}$ 
commutes with $x$ and $p$ since it does not contain $x$ and $\partial_x$.

We consider the operation of the Heisenberg position operator $x(t)-1$ to the wavepacket
(We consider $x(t)-1$ instead of $x(t)$ for the simplicity of the following calculations.): 
\begin{equation}
 \begin{split}
  &[x(t)-1]\, \phi(x)=U^\dag(t)\, (x-1)\, U(t)\phi(x)\\
  &\simeq U^\dag(t)\, (x-1) \left[\phi_\textrm{free}(t,x_+)-\phi_\textrm{free}(t,x_-)
  \right]h(x)\ .
 \end{split}
\end{equation}
At the last equality, we used Eq.(\ref{Nbounce}). 
By the similar way as the momentum operator,
the operation of $(x-1)$ to free wavepacket can be written by using $k_0$-derivative as
\begin{equation}
(x-1)\phi_\textrm{free}(t,x_\pm)=\pm \mathcal{B}\phi_\textrm{free}(t,x_\pm)\ ,
\end{equation}
where
\begin{equation}
 \mathcal{B}\equiv -i\alpha(t)\partial_{k_0}+\ell(t)\ ,\quad \ell(t)\equiv 2k_0 t+x_0-2N-1\ .
\end{equation}
The operator $\mathcal{B}$ commutes with $x$ and $p$. 
The introduced variable $\ell(t)$ represents the coordinate difference between the center of the free wavepacket $x=x_\textrm{center}$
and the position of $(2N+1)$-th bounce $x=2N+1$.
One can check that the introduced operators $\mathcal{A}$ and $\mathcal{B}$ satisfy the ``canonical commutation relation'':
\begin{equation}
 [\mathcal{A},\mathcal{B}]=i\ .
  \label{ABcom}
\end{equation}
Using the operator $\mathcal{B}$, we obtain 
\begin{equation}
 [x(t)-1]\, \phi(x)=\mathcal{B} U^\dag(t) \left[\phi_\textrm{free}(t,x_+)+\phi_\textrm{free}(t,x_-)
		\right]h(x)\ .
 \label{xphi1}
\end{equation}
We need to calculate the inverse time evolution of $\phi_\textrm{free}(t,x_\pm)h(x)$.
The strategy is same as the previous subsection:
We consider the inverse time evolution by the free Hamiltonian $U_\textrm{free}^\dag$ and apply the folding operator $F$.
The propagator of the free particle is given by
\begin{equation}
 K(x,t;x',t_0)=\frac{1}{\sqrt{4\pi i(t-t_0)}}\exp\left[
\frac{i(x-x')^2}{4(t-t_0)}
					     \right]\ .
\end{equation}
Using the propagator, we have
\begin{equation}
 \begin{split}
  &U_\textrm{free}^\dag(t)\phi_\textrm{free}(t,x_\pm)h(x)
  =\int^{1}_0 K(x,0;x',t)\phi_\textrm{free}(t,x'_\pm)\\
  &\simeq \int^{1}_{-\infty} K(x,0;x',t)\phi_\textrm{free}(t,x'_\pm)\ ,
 \end{split}
\end{equation}
where $x'_+=x'+2N$ and $x'_-=2N+2-x'$.
At the last equality, we extended the lower bound of the integration to $-\infty$ because
we assumed that the wavepacket is not around the left boundary.
Completing the square in the exponent of the integrand, we can rewrite above expression as
\begin{equation}
  \frac{1}{(2\pi\sigma^2)^{1/4}}\sqrt{\frac{a}{\pi}}  \int^{1}_{-\infty} \exp\left[
-a\{x'-1\mp \xi(x_\pm)\}^2+b(x_\pm)
	    \right]\ ,
 \label{Kphi}
\end{equation}
where
\begin{equation}
\begin{split}
&a=-\frac{1}{4it\alpha(t)}\ ,\qquad
b(x)= -\frac{x-x_0}{4\sigma^2}+ik_0 x\ ,\\
&\xi(x)=\alpha(t)(x-x_0)+\ell(t)\ .
\end{split}
\end{equation}
Note that we can rewrite $(2\pi\sigma^2)^{-1/4}e^{b(x_\pm)}=\phi(x_\pm)$ in Eq.(\ref{Kphi}).
Using the error function\footnote{
We define the error functions for $z\in \bm{C}$ as
\begin{equation}
 \textrm{erf}(z)=\frac{2}{\sqrt{\pi}}\int^{z}_0dz e^{-z^2}\ ,\quad
  \textrm{erfc}(z)=1-\textrm{erf}(z)\ .
\end{equation}
}, we have
\begin{equation}
 U_\textrm{free}^\dag(t)\phi_\textrm{free}(t,x_\pm)h(x)=\frac{1}{2}\phi(x_\pm)\textrm{erfc}[\pm \sqrt{a}\xi(x_\pm)]\ .
\end{equation}
Taking the folding operation, we obtain inverse time evolution of the wave function in a box as
\begin{equation}
\begin{split}
 &U^\dag(t)\phi_\textrm{free}(t,x_\pm)h(x)
 =\frac{1}{2}F\left\{\phi(x_\pm)\textrm{erfc}[\pm \sqrt{a}\xi(x_\pm)]\right\}\\
 =&\frac{1}{2}F\left\{\phi(\pm x)\textrm{erfc}[\pm \sqrt{a}\xi(\pm x)]\right\}
 =\pm \frac{1}{2}F\left\{\phi(x)\textrm{erfc}[\pm \sqrt{a}\xi(x)]\right\}
\end{split}
\end{equation}
At the second equality, we used the definition of $x_\pm$~(\ref{xpmdef}) and the formula of folding operator~(\ref{Fform1}).
At the last equality, we used the other formula~(\ref{Fform2}). 
So, we obtain
\begin{equation}
 U^\dag(t)[\phi_\textrm{free}(t,x_+)+\phi_\textrm{free}(t,x_-)]h(x)=F\{\Phi(x)\}
 \label{Uphi}
\end{equation}
where we used $\textrm{erf}(-z)=-\textrm{erf}(z)$ and defined 
\begin{equation}
 \Phi(x)\equiv -\phi(x)\textrm{erf}[\sqrt{a}\xi(x)]\ .
\end{equation}
From Eq.(\ref{xphi1}), the operation of $x(t)-1$ to the gaussian wavepacket is given by
\begin{equation}
 [x(t)-1]\,\phi(x)=\mathcal{B}F\{\Phi(x)\}\ .
  \label{xphi2}
\end{equation}

\subsection{2-point out-of-time-order correlator}

We can easily obtain analytic expression for the 2-point OTOC.
The 2-point OTOC is given by
\begin{equation}
\begin{split}
 b_\phi&=-i\langle \phi|[x(t),p]|\phi\rangle = -2\,\textrm{Im}\,\langle \phi|\,[x(t)-1]\,p\,|\phi\rangle\\
 &=-2\,\textrm{Im}\,\langle \phi|\,U^\dag(t)\,(x-1)\,U(t)\,\mathcal{A}|\phi\rangle
 =-2\,\textrm{Im}\,\langle \phi(t)|\,(x-1)\,\mathcal{A}\,|\phi(t)\rangle\ .
\end{split}
\end{equation}
At the third equality, we replaced momentum operator by $\mathcal{A}$.
We already know the wave function $|\phi(t)\rangle $ as in Eq.(\ref{Nbounce}).
Therefore, the 2-point OTOC is written as
\begin{equation}
 b_\phi
  \simeq
  -2\textrm{Im} \int^1_{-\infty}dx\, (x-1) [\phi_\textrm{free}^\ast(t,x_+)\mathcal{A}\phi_\textrm{free}(t,x_+)+\phi_\textrm{free}^\ast(t,x_-)\mathcal{A}\phi_\textrm{free}(t,x_-)]\ .
\end{equation}
We neglected the cross terms
such as $\phi^\ast(t,x_+)\phi(t,x_-)$ since they oscillate very quickly as $\sim e^{\pm 2ik_0 x}$ and canceled out by the integration.
Substituting the explicit expression of $\phi_\textrm{fee}(t,x)$~(\ref{phifreet}) and introducing $x'=x-1$, we obtain the 2-point OTOC as
\begin{equation}
\begin{split}
 b_\phi
  &=
  -\frac{1}{(2\pi)^{1/2}\sigma_t^3} \int^0_{-\infty}dx' x'
 [(x'-\ell)e^{-\frac{(x'-\ell)^2}{2\sigma_t^2}}-(x'+\ell)e^{-\frac{(x'+\ell)^2}{2\sigma_t^2}}]\\
 &=\textrm{erf}\left[\frac{\ell(t)}{\sqrt{2}\sigma_t}\right]\ .
\end{split}
\end{equation}
At the bounce, the 2-point OTOC changes the signature. Its time scale is given by
$\Delta t\sim \sigma_t/k_0$. 
This is consistent with the numerical calculation in Fig.\ref{bphifig}.

\subsection{Operation of $[x(t),p(0)]$ to the wavepacket}

For the analytic calculation of 4-point OTOC,
 we consider the operation of $[x(t),p(0)]$ to the wavepacket.
From Eqs.(\ref{Aphi}) and (\ref{xphi2}), we obtain
\begin{equation}
 \begin{split}
 &[x(t)-1]\,p\, \phi(x)=\mathcal{A}\,\mathcal{B}\,F\{\Phi(x)\}\ ,\\
 &p\,[x(t)-1]\,\phi(x)=\mathcal{B}\,p\,F\{\Phi(x)\}\ .
 \end{split}
\end{equation}
Note that, in the second line, we cannot replace the momentum operator $p$ by $\mathcal{A}$
since it is not applied to the initial gaussian wavepacket.
Thus, the operation of $[x(t),p(0)]$ to the gaussian wavepacket is given by
\begin{equation}
\begin{split}
 \Psi(x)&\equiv -i[x(t),p(0)]\phi(x)\\
 &=-i\{[\mathcal{A},\mathcal{B}]+\mathcal{B}(\mathcal{A}-p)\}F\{\Phi(x)\}\\
 &=[1 -i\mathcal{B}(\mathcal{A}-p)]F\{\Phi(x)\}\ .
\end{split}
\end{equation}
At the last equality, we used Eq.(\ref{ABcom}).
Now, we consider $-i\mathcal{B}(\mathcal{A}-p)F\{\Phi(x)\}$. Here, $F\{\Phi(x)\}$ is composed of right moving part $\Phi(x+2m)$ and left moving part $\Phi(2m-x)$. 
By a explicit calculation, we can check that the right moving contribution is zero:
\begin{equation}
 \mathcal{B}(\mathcal{A}-p)\Phi(x+2m)=0\ .
\end{equation}
The left moving contribution is given by
\begin{equation}
 \begin{split}
 -\frac{1}{2}\mathcal{B}(\mathcal{A}-p)\Phi(2m-x)&=
  \left\{\left(k_0+i\frac{y-x_0}{2\sigma^2}\right)\xi(y)-i\alpha(t)\right\}\Phi(y)\\
 &\hspace{3cm}+\frac{1}{\pi a}\left(k_0+i\frac{y-x_0}{2\sigma^2}\right)e^{-a\xi(y)^2}\phi(y)\ .
\end{split}
\end{equation}
where $y=2m-x$ represents the coordinate before the folding operation. Then, $\Psi(x)$ is written as 
\begin{equation}
 \Psi(x)= \Psi_R(x)+\Psi_L(x)\ ,
\end{equation}
where $\Psi_R$ and $\Psi_L$ represent right and left moving contributions: 
\begin{equation}
 \Psi_R(x)=\sum_m \Phi(y)\big|_{y=x+2m}\ ,
  \label{PhiR1}
\end{equation}
and
\begin{multline}
\Psi_L(x)=-2i\sum_m \bigg[\left\{\left(k_0+i\frac{y-x_0}{2\sigma^2}\right)\xi(y)-i\alpha(t)-\frac{i}{2}\right\}\Phi(y)\\
+\frac{1}{\sqrt{\pi a}}\left(k_0+i\frac{y-x_0}{2\sigma^2}\right)e^{-a\xi(y)^2}\phi(y)\bigg]_{y=2m-x}\ . 
\end{multline}
In the curly bracket of $\Psi_L$, $-i\alpha(t)-i/2$ is negligible.
We can see that as follows.
In the expression of $k_0 \xi(y)$,
there is a term of $k_0 \alpha(t)(y-x_0)$.
Here, $y=2m-x$ is outside the region of the box: $y<0$ or $y>1$.
Thus, from Eq.(\ref{x0cond}), we have
\begin{equation}
 |y-x_0|\gg \sigma\ .
  \label{ycond}
\end{equation}
It follows $k_0|y-x_0|\gg 1$ from well localized condition~(\ref{local}).
Therefore, we obtain a relation: $|k_0 \alpha(t)(y-x_0)|\gg |i\alpha(t)| > |i/2|$.
As the result, we can rewrite $\Psi_L$ as
\begin{equation}
 \Psi_L(x)\simeq -\frac{2i}{\sqrt{a}}\sum_m \left(k_0+i\frac{y-x_0}{2\sigma^2}\right)\textrm{Ierf}[\sqrt{a}\xi(y)]\phi(y)\,\bigg|_{y=2m-x}\ ,
\label{PsiL1}
\end{equation}
where
\begin{equation}
 \textrm{Ierf}(z)\equiv z\textrm{erf}(z)+\frac{1}{\sqrt{\pi}}e^{-z^2}\ ,
\end{equation}
is the primitive integral of $\textrm{erf}(z)$: $d\textrm{Ierf}(z)/dz=\textrm{erf}(z)$.

The expression of $\Psi_L$ is still complicated. To obtain a simpler expression, we take the relatively-late-time-approximation:
\begin{equation}
 \sigma^2\ll t \ll \sigma\ .
\end{equation}
The upper inequality is from Eq.(\ref{tE}).
The lower inequality implies that the width of the wavepacket is much wider than the initial width ($\sigma_t \gg \sigma$). Expanding $\sqrt{a}\xi(y)$ in terms of $\sigma^2/t$, we have
\begin{equation}
 \sqrt{a}\xi(y)\simeq iq
 +\frac{1}{2}\left(q+\frac{\ell}{\sigma}\right)\frac{\sigma^2}{t}
 +\frac{i}{8}\left(q+\frac{2\ell}{\sigma}\right)\left(\frac{\sigma^2}{t}\right)^2+\cdots\ .
\end{equation}
where
\begin{equation}
 q\equiv \frac{y-x_0}{2\sigma}\ .
\end{equation}
From Eq.(\ref{ycond}), we obtain $|q|\gg 1$. 
So, we only need the asymptotic form of the $\textrm{Ierf}[\sqrt{a}\xi(y)]$.
For large $|z|$, the asymptotic expression of the error function is 
\begin{equation}
 \textrm{erfc}(z)\sim \frac{e^{-z^2}}{\sqrt{\pi}z}(1-\frac{1}{2z^2}+\cdots)\ .
\end{equation}
Using this expansion, we have
\begin{equation}
\textrm{Ierf}[\sqrt{a}\xi(y)]\simeq -\frac{1}{2\sqrt{\pi}q^2}
  \exp\left[
  q^2
  -iq\left(q+\frac{\ell}{\sigma}\right)\frac{\sigma^2}{t}
  -\frac{\ell^2}{4\sigma^2}\left(\frac{\sigma^2}{t}\right)^2
  \right]\ ,
\end{equation}
where we considered up to second order of $\sigma^2/t$ in the exponent.
On the other hand, we only took into account
the leading term outside the exponential.
Using above expression, we can rewrite the left moving contribution as
\begin{multline}
 \Psi_L(x)\simeq \frac{2\sqrt{2}it}{(2\pi\sigma^2)^{3/4}}\exp\left(-\frac{\sigma^2\ell^2}{4t^2}\right)
   \sum_m \left(k_0+i\frac{q}{\sigma}\right)\\
  \times \frac{1}{q^2}
  \exp\left[iq\sigma\left\{2k_0-\left(q+\frac{\ell}{\sigma}\right)\frac{\sigma}{t}\right\}\right]\,
 \bigg|_{q=(2m-x-x_0)/(2\sigma)}\ ,
\label{PsiL}
\end{multline}
Similarly, in the expression of $\Psi_R$~(\ref{PhiR1}), for $m\neq 0, $we can replace $\textrm{erf}[\sqrt{a}\xi(y)]$ by the asymptotic expression  as 
\begin{equation}
 \textrm{erf}[\sqrt{a}\xi(y)]\simeq \frac{i}{\sqrt{\pi}q}
  \exp\left[
  q^2
  -iq\left(q+\frac{\ell}{\sigma}\right)\frac{\sigma^2}{t}
  -(\frac{\sigma\ell}{2t})^2
  \right]\bigg|_{q=(x-2m-x_0)/(2\sigma)}\ .
\end{equation}
Note that, for $m=0$, $q=(x-x_0)/(2\sigma)$ can be small. So,  above expression is not available.
We can see that contributions from $m\neq 0$ are suppressed by $1/q$ compared to $m=0$.
Therefore, the main contribution for $\Psi_R$ is $m=0$:
\begin{equation}
 \Psi_R(x)\simeq \Phi(x)\ .
\end{equation}

\subsection{4-point Out-of-time-order correlator}

The 4-point OTOC is given by
\begin{equation}
\begin{split}
 c_\phi&=-\int^1_0 dx \phi^\ast(x)[x(t),p(0)]^2\phi(x)=\int^1_0 dx \Psi^\ast(x) \Psi(x)\\
 &\simeq\int^1_0 dx [\Psi_R^\ast(x) \Psi_R(x)+\Psi_L^\ast(x) \Psi_L(x)]\ .
\end{split}
\end{equation}
The cross terms of right and left movers are negligible in the integration.
Recall that
\begin{equation}
 \Psi_R(x)\simeq \Phi(x) \simeq F\{\Phi(x)\}=U^\dag(t)\left[\phi_\textrm{free}(t,x_+)+\phi_\textrm{free}(t,x_-)
		\right]h(x)\ .
\end{equation}
At the second equality, we used the fact that $\Phi(y)$ is suppressed by $1/q=2\sigma/(y-x_0)\ll 1$ outside the domain of the box.
At the third equality, we used Eq.(\ref{Uphi}).
The right moving contribution to $c_\phi$ is given by
\begin{equation}
 \int^1_0 dx \Psi_R^\ast(x) \Psi_R(x)=\int^1_0 dx \{|\phi(x_+)|^2+|\phi(x_-)|^2\}=1\ .
\end{equation}
Let us consider the left moving contribution $\Psi_L^\ast \Psi_L$.
As in Eq.(\ref{PsiL}), $\Psi_L$ is written in the form of $\sum_{m}(\cdots)$.
So, $\Psi_L^\ast \Psi_L$ is written as $\sum_{m,n}(\cdots)$. 
We focus on its cross term of $n$ and $m$. 
Its exponent is given by 
\begin{multline}
 \left[iq\sigma\left\{2k_0-\left(q+\frac{\ell}{\sigma}\right)\frac{\sigma}{t}\right\}\right]_{q=\frac{2m-x-x_0}{2\sigma}}
 -\left[iq\sigma\left\{2k_0-\left(q+\frac{\ell}{\sigma}\right)\frac{\sigma}{t}\right\}\right]_{q=\frac{2n-x-x_0}{2\sigma}}\\
 =\frac{i}{t}(m-n)(x+x_0+n-m-1)
\end{multline}
From Eqs.(\ref{local}) and (\ref{tE}), we have $1/t=(1/\sigma) (\sigma/t)\gg 1$.
So, the exponent is quickly rotating for $m\neq n$ and the cross terms are negligible in the integral.
Therefore, left moving contribution to $c_\phi$ is simply given by
\begin{equation}
 \int^1_0 dx \Psi_L^\ast(x) \Psi_L(x)\simeq
    \frac{8t^2}{(2\pi\sigma^2)^{3/2}}e^{-\frac{\sigma^2\ell^2}{2t^2}}
   \sum_m \int^1_0 dx \left(k_0^2+\frac{q^2}{\sigma^2}\right)\frac{1}{q^4}\bigg|_{q=\frac{2m-x-x_0}{2\sigma}}
\end{equation}
We can perform the integral and the summation over $m$ as
\begin{equation}
\begin{split}
 \sum_{m=-\infty}^\infty\int_0^1 dx \frac{1}{q^4}\bigg|_{q=\frac{2m-x-x_0}{2\sigma}}
  &=\frac{16\sigma^4}{3}\sum_{m=-\infty}^\infty \frac{3(2m-x_0)^2-3(2m-x_0)+1}{(2m-x_0)^3(2m-x_0-1)^3}\\
 &=\frac{8\pi^3\sigma^4}{3}\frac{1+\cos^2\pi x_0}{\sin^3\pi x_0}\ .
\end{split}
\end{equation}
and
\begin{equation}
 \sum_{m=-\infty}^\infty\int_0^1 dx \frac{1}{q^2}\bigg|_{q=\frac{2m-x-x_0}{2\sigma}}
  =\sum_{m=-\infty}^\infty\frac{4\sigma^2}{(2m-x_0)(2m-x_0-1)}
  =\frac{4\pi\sigma^2}{\sin\pi x_0}\ .
\end{equation}
Therefore, we obtain
\begin{equation}
 \begin{split}
 c_\phi&=
  1+\left(\frac{128}{\pi}\right)^{1/2}
  \left\{\frac{2\pi^3 \sigma k_0^2(1+\cos^2\pi x_0)}{3\sin^3\pi x_0}+\frac{1}{\sigma^3\sin\pi x_0}\right\}t^2 e^{-\frac{\sigma^2\ell^2}{2t^2}}\\
  &\simeq
  1+
  \left\{\frac{42.0 \sigma k_0^2(1+\cos^2\pi x_0)}{\sin^3\pi x_0}+\frac{6.38}{\sigma^3\sin\pi x_0}\right\}t^2 e^{-\frac{\sigma^2\ell^2}{2t^2}}\ .
 \end{split}
\end{equation}
In addition to the classical prediction of the 4-point OTOC, $c_\phi^\textrm{classical}=1$,
we find the gaussian spike at the bounce.


\begin{thebibliography}{99}

\bibitem{Larkin}
A.~I.~Larkin and Y.~N.~Ovchinnikov,
JETP 28, 6 (1969): 1200-1205.


\bibitem{Maldacena:2015waa} 
  J.~Maldacena, S.~H.~Shenker, and D.~Stanford,
  JHEP {\bf 1608}, 106 (2016)
  [arXiv:1503.01409 [hep-th]].
  
\bibitem{Maldacena:1997re} 
  J.~M.~Maldacena,
  Int.\ J.\ Theor.\ Phys.\  {\bf 38}, 1113 (1999)
  [Adv.\ Theor.\ Math.\ Phys.\  {\bf 2}, 231 (1998)]
  [hep-th/9711200].

  
\bibitem{Shenker:2013pqa} 
  S.~H.~Shenker and D.~Stanford,
  JHEP {\bf 1403}, 067 (2014)
  [arXiv:1306.0622 [hep-th]].


\bibitem{Shenker:2013yza} 
  S.~H.~Shenker and D.~Stanford,
  JHEP {\bf 1412}, 046 (2014)
  [arXiv:1312.3296 [hep-th]].
  
\bibitem{Leichenauer:2014nxa} 
  S.~Leichenauer,
  Phys.\ Rev.\ D {\bf 90}, no. 4, 046009 (2014)
  [arXiv:1405.7365 [hep-th]].

\bibitem{Kitaev-talk}
A.~Kitaev, 
talk given at Fundamental Physics Symposium, Nov.~2014.


\bibitem{Shenker:2014cwa} 
  S.~H.~Shenker and D.~Stanford,
  JHEP {\bf 1505}, 132 (2015)
  [arXiv:1412.6087 [hep-th]].
  
\bibitem{Jackson:2014nla} 
  S.~Jackson, L.~McGough, and H.~Verlinde,
  Nucl.\ Phys.\ B {\bf 901}, 382 (2015)
  [arXiv:1412.5205 [hep-th]].
  

\bibitem{Polchinski:2015cea} 
  J.~Polchinski,
  arXiv:1505.08108 [hep-th].




\bibitem{Sachdev:1992fk} 
  S.~Sachdev and J.~Ye,
  Phys.\ Rev.\ Lett.\  {\bf 70}, 3339 (1993)
  [cond-mat/9212030].

\bibitem{Kitaev-talk-KITP}
A. Kitaev, 
talks given at KITP, April and May 2015.


\bibitem{Gross}
D.\ J.\ Gross and V.\ Rosenhaus, 
arXiv:1610.01569v1 (2016). 
\bibitem{Witten}
E.\ Witten, 
arXiv:1610.09758v2 (2016). 
\bibitem{Terashima}
T.\ Nishinaka and S.\ Terashima, 
arXiv:1611.10290v1 (2016). 




\bibitem{Sinai}
Ya.~G.~Sinai,
Russ. Math. Surv. {\bf 25}, 137 (1970).

\bibitem{Bunimovich1}
L.~A.~Bunimovich, Funct. Anal. Appl. {\bf 8}, 254 (1974).
	 
\bibitem{Bunimovich2}
L.~A.~Bunimovich, Commun. Math. Phys. {\bf 65}, 295 (1979).

\bibitem{Bunimovich3}
L.~A.~Bunimovich, Ya.~G.~Sinai, and N.~J.~Chernov,
	Russ. Math. Surv. {\bf 46}, 47 (1991).
	




\bibitem{Benettin}
G.~Benettin and J.~M.~Strelcyn, Phys. Rev. A {\bf 17}, 773 (1978).



\bibitem{Hashimoto:2016wme} 
  K.~Hashimoto, K.~Murata, and K.~Yoshida,
  Phys.\ Rev.\ Lett.\  {\bf 117}, no. 23, 231602 (2016)
  [arXiv:1605.08124 [hep-th]].


	
\bibitem{Dellago}
C.~Dellago and H.~A.~Posch, Phys. Rev. E {\bf 52}, 3, 2401 (1995).

\bibitem{Biham}
O.~Biham and M.~Kvale, Phys. Rev. A {\bf 46}, 6334 (1992).

\bibitem{McDonald}
S.~W.~McDonald. A.~N.~Kaufman,
Phys.~Rev.~Lett.~{\bf 42}, 18 (1979).

\bibitem{Berry}
M.~Berry, 
``Chaos and the semiclassical limit of quantum mechanics (is the moon there when somebody looks?)``,
in Quantum mechanics: Scientfic perpectives on Divine Action, pp41-54.
	
\bibitem{Zurek1}
W.~H.~Zurek and J.~P.~Paz, 
	Phys.~Rev.~Lett. {\bf 72}, 2508-2511 (1994),
\bibitem{Zurek2}
	W.~H.~Zurek and J.~P.~Paz, 
	PHYSICA D {\bf 83}, 300-308 (1995),
\bibitem{Zurek3}
	W.~H.~Zurek,
	 Physica Scripta {\bf 76}, 186-198 (1998). 

\bibitem{Bhattacharya}
T.~Bhattacharya, S.~Habib, and K.~Jacobs, 
Phys.~Rev.~Lett.~ {\bf 85}, 4852 (2000). 

 \bibitem{Rozenbaum}
E.~B.~Rozenbaum, S.~Ganeshan, and V.~Galitski, 
  Phys.\ Rev.\ Lett.\  {\bf 118}, 086801 (2017)
  arXiv:1609.01707 [cond-mat.dis-nn].

\bibitem{Prosen}
T.~Prosen, T.~H.~Seligman, and M.~Znidaric,
Prog.~Theo.~Phys.~Supp. {\bf 150}, 200 (2003)
arXiv:quant-ph/0304104.
	 








\bibitem{Aleiner}
I.~L.~Aleiner, L.~Faoro, and L.~B.~Ioffe,
Annals of Physics {\bf 375}, 378 (2016)
arXiv:1609.01251 [cond-mat.stat-mech].

\bibitem{Campisi}
M.~Campisi and J.~Goold,
arXiv:1609.05848 [quant-ph]. 

\bibitem{Tsuji}
N.~Tsuji, P.~Werner, and M.~Ueda,
Phys.~Rev. {\bf A 95}, 011601(R) (2017)
arXiv:1610.01251 [cond-mat.quant-gas].


\bibitem{Kurchan}
J.~Kurchan, 
arXiv:1612.01278 [cond-mat.stat-mech]. 

\bibitem{Bohrdt}
A.~Bohrdt, C.~B.~Mendl, M.~Endres, and M.~Knap,
arXiv:1612.02434 [cond-mat.quant-gas].

\bibitem{Tsuji2}
N.~Tsuji, T.Shihata, and M.~Ueda,
arXiv:1612.08781 [cond-mat.stat-mech]. 

\bibitem{Kuku}
I.~Kukuljan, S.~Grozdanov, and T.~Prosen, 
	arXiv:1701.09147 [cond-mat.stat-mech].
	
\bibitem{Zeh}
    H.~D.~Zeh, 
     Phys. Lett. A {\bf 116}, 9 (1986).

\bibitem{Kiefer}
    C.~Kiefer, 
     Phys.~Rev.~D {\bf 46}, 1658 (1992). 

\bibitem{Adachi} 
    S.~Adachi, M.~Toda, and K.~Ikeda, 
     Phys.~Rev.~Lett.~{\bf 61}, 659 (1988). 

\bibitem{Castagnino}
     M.~Castagnino, S.~Fortin, and O.~Lombardi, 
      Mod.~Phys.~Lett.~A {\bf 25}, 611 (2010). 
     
\bibitem{Fishman} 
    S.~Fishman, D.~R.~Grempel and R.~E.~Prange, 
	Phys.~Rev.~Lett.~{\bf 45}, 509 (1982).
\end{thebibliography}
\end{document}